\documentclass[lettersize,journal]{IEEEtran}
\usepackage{amsmath,amsfonts}
\usepackage{amssymb}
\usepackage{array}
\usepackage{textcomp}
\usepackage{stfloats}
\usepackage{url}
\usepackage{verbatim}
\usepackage{graphicx}
\usepackage{amsthm}
\usepackage{cite}
\usepackage{soul, color, xcolor}
\usepackage{algorithm}  
\usepackage{algpseudocode}  
\usepackage{subfigure}
\usepackage{multirow}

\usepackage{amsmath}  

\makeatletter
\renewcommand{\maketag@@@}[1]{\hbox{\m@th\normalsize\normalfont#1}}%
\makeatother
\begin{document}

\title{Security Defense of Large-scale Networks Under False Data Injection Attacks: An Attack Detection Scheduling Approach}
\author{Yuhan Suo, Senchun Chai,~\IEEEmembership{Senior Member,~IEEE,} Runqi Chai,~\IEEEmembership{Member,~IEEE,} \\  Zhong-Hua Pang,~\IEEEmembership{Senior Member,~IEEE,} Yuanqing Xia,~\IEEEmembership{Fellow,~IEEE,} and Guo-Ping Liu,~\IEEEmembership{Fellow,~IEEE.}
        \thanks{This work was supported by the Open Research Fund Program of Key Laboratory of Industrial Internet and Big Data under Grant IIBD-2021-KF01.(\emph{Corresponding author: Senchun Chai})}
\thanks{Yuhan Suo, Senchun Chai, Runqi Chai, and Yuanqing Xia are with the School of Automation, Beijing Institute of Technology, Beijing 100081, China (e-mail: yuhan.suo@bit.edu.cn; chaisc97@bit.edu.cn; r.chai@bit.edu.cn; xia$\_$yuanqing@bit.edu.cn). }

\thanks{Zhong-Hua Pang is with the Key Laboratory of Fieldbus Technology
	and Automation of Beijing, North China University of Technology,
	Beijing 100144, China (e-mail:
	zhonghua.pang@ia.ac.cn).}

\thanks{Guo-Ping Liu is with the Department of Artificial Intelligence,
	and Automation, Wuhan University, Wuhan 430072, China (e-mail:
	guoping.liu@southwales.ac.uk).}
}
\markboth{IEEE TRANSACTIONS ON  INFORMATION FORENSICS AND SECURITY}%
{Shell \MakeLowercase{\textit{et al.}}: A Sample Article Using IEEEtran.cls for IEEE Journals}

\maketitle

\begin{abstract}

In large-scale networks, communication links between nodes are easily injected with false data by adversaries. This paper proposes a novel security defense strategy from the perspective of attack detection scheduling to ensure the security of the network. Based on the proposed strategy, each sensor can directly exclude suspicious sensors from its neighboring set. 
First, the problem of selecting suspicious sensors is formulated as a combinatorial optimization problem, which is non-deterministic polynomial-time hard (NP-hard). To solve this problem, the original function is transformed into a submodular function. Then, we propose an attack detection scheduling  algorithm based on the sequential submodular optimization theory, which incorporates \emph{expert problem} to better utilize historical information to guide the sensor selection task at the current moment.
For different attack strategies, theoretical results show that the average optimization rate of the proposed algorithm has a lower bound, and the error expectation is bounded.
In addition, under two kinds of insecurity conditions, the proposed algorithm can guarantee the security of the entire network from the perspective of the augmented estimation error. Finally, the effectiveness of the developed method is verified by the numerical simulation and practical experiment.

\end{abstract}

\begin{IEEEkeywords}
Networks security, attack detection scheduling, sequential submodular optimization, secure state estimation, security analysis.
\end{IEEEkeywords}

\section{Introduction}
In recent years, with the advancement of network technology, nodes in large-scale networks are able to communicate in real-time and collaborate to complete complicated tasks. However, malicious attackers attempt to compromise the network's security by attacking communication links between nodes \cite{pang2022security}.  
\begin{itemize}
	\item{In a cooperative ground-air system, where multiple UAVs need real-time communication to share target location and jointly track the ground targets. However, attackers can interfere with the accuracy of target location information by injecting false signals \cite{jia2017modelling}.}
	\item{In a distributed power system, multiple sensors jointly monitor the state of the system, and attackers can also prevent some sensors from accurately estimating the state of the system
\cite{jin2020boundary}.}
	\item{In Connected and Automated Vehicles (CAVs) systems, malicious vehicles can spread false information, which affects the decision-making of surrounding vehicles and threatens people's lives and properties
 \cite{yang2021risk}.}
\end{itemize}

Due to the difficulty of supplying constant power to distributed nodes, as well as the limitations of physical size and cost, the energy and computing power of each edge node in the above distributed large-scale networks are limited \cite{segura2021centralized}. As a result, an efficient security defense strategy that can not only minimize the energy consumption and computing power requirement of each edge node but also avoid the impact of malicious information on network security is required.

In recent years, the security issues of large-scale networks under false data injection attacks (FDIAs) have been widely studied, and attack detection algorithms and resilient defense mechanisms are considered to be effective against FDIAs. Effective attack detection algorithms can help the system to detect malicious adversaries in real-time, which can be divided into the following four categories:

The $\chi^2$ detector is a common residual-based attack detector that has been widely investigated \cite{brumback1987chi}. However, well-crafted attack signals can bypass the $\chi^2$ detector and threaten the security of the network \cite{zhou2022security,liu2022completely}.
However, the covariance matrix of the $\chi^2$ detector must be invertible, which is difficult to implement in practice \cite{chen2021resilient,lv2022resilient}. Indeed, it is difficult for $\chi^2$ detectors with a fixed detection threshold to detect well-crafted attack signals \cite{pang2023analysis}. Therefore, Han et al. \cite{han2015stochastic} and Zhou et al. \cite{zhou2022security} carried out research on the design of the detection threshold.
To take full advantage of the internal connectivity of distributed networks to aid attack detection, Ferrari et al. \cite{ferrari2011distributed} studies the problem of fault detection and isolation in the case of coupling between subsystems. 
To deal with the problem that a single node cannot obtain the global information of the entire distributed network, Ge et al. \cite{ge2019distributed} and Ju et al. \cite{ju2020distributed} both designed a distributed estimators to estimate the system state using local information. On this basis, effective attack detection algorithms were designed respectively based on the residuals obtained by the designed distributed estimator.

The representative active detection approach is the watermark-based attack detection approach proposed by Mo et al. \cite{mo2009secure}. On this basis, Yang et al. \cite{yang2019distributed} detected man-in-the-middle attacks by detecting extra verification data attached to transmitted packets.  Although this type of attack detection approach is effective, the additional control cost is not friendly to edge nodes with limited energy and computing power in distributed large-scale networks.
Recently, \cite{fang2020optimal,pang2021detection,zhou2022secure} investigated the trade-off between the control cost and detection performance of attack detection approaches based on the watermarking mechanism. In addition, Xu et al. \cite{xu2022robust} investigated the design of robust moving target defense in power grids, which limits the chance of undetectable subspaces being attacked.

Reachability analysis was used to analyze the impact of attacks on network control systems in the early stage \cite{kwon2013security}, and Mousavinejad te al. \cite{mousavinejad2018novel} applied it to the field of attack detection. Li et al. \cite{li2023attack} proposed an attack detection approach based on the partition reachability analysis, which detects attacks based on the intersection between the predicted state set and the measured state set. In fact, to obtain better detection performance, a significant quantity of computational power is required (to obtain a tighter set).

With the improvement of computing power, data-driven attack detection methods have gradually emerged.
Li et al. \cite{li2019data} proposed an attack detection method based on data-driven and hybrid optimization strategies to deal with sparse attacks in large scale networks with unknown dynamic characteristics. 
And Liu et al. \cite{liu2021flipit} calculated the probability distribution of attack system compromise time based on the data obtained by Monte Carlo simulation (MCS).
Using the subspace approach, Zhao et al. \cite{zhao2022data} proposed a data-driven attack detection strategy and attack identification scheme.  
To deal with unknown attacks in distributed power grids, Peng et al. \cite{peng2023localizing} proposed a detection and localization method based on neural networks.
However, how to collect comprehensive and credible data to improve data-driven attack detection methods in large-scale networks is challenging.

Different from the aforementioned attack detection algorithms, the resilient defense mechanism, also known as secure state estimation,  can ensure that the system obtains accurate state estimates in the presence of malicious attacks by enhancing the fault tolerance of the system itself.
In the case that the information of some nodes (sensors) is maliciously tampered with, the secure state estimates of the system can be obtained based on redundant information \cite{shoukry2017secure}.
However, some existing literature considers that the problem of obtaining an unknown set of attacked sensors is NP-hard
\cite{lu2023polynomial,an2022fast,lu2023secure}. To avoid the combinatorial explosion caused by brute-force search, Lu et al. \cite{lu2023polynomial} reconstructed the system state based on the majority voting, and An et al. \cite{an2022fast} designed a fast state estimation algorithm considering the equivalence between the measurements of the sensors.
In addition, sensor fusion algorithms can also be used to obtain secure state estimates. Yang et al. \cite{yang2021secure} proposed a sensor fusion algorithm based on information redundancy, which improves the resilience of the CAVs to malicious vehicles. For nonlinear systems under FDIAs, Weng et al. \cite{weng2023secure} proposed a learning-based local information fusion method to minimize the estimation error of the system.

It can be seen from the existing work that there are few studies on the problem of attack detection in distributed large-scale networks, and the current research generally detects information from each neighboring sensor, which usually has high energy and computing power requirement. Therefore, we are motivated to consider  whether the cost of attack detection  can be reduced by directly searching the set of suspicious sensors, 
that is, it is no longer necessary to detect the information of all neighboring sensors. The key of the problem is how to select suspicious sensors (note that this is a combinatorial optimization problem, which is NP-hard and challenging). In addition, the existing literature usually studies the \emph{static attack strategy} because the \emph{dynamic attack strategy} is more complicated \cite{liu2021flipit}. Therefore, we also expect that the proposed security defense strategy can cope with the \emph{dynamic attack strategy}, which will greatly enhance its flexibility. The main contributions of this paper are as follows:

\begin{itemize}
	\item{This paper proposes a novel security defense strategy from the perspective of attack detection scheduling to ensure the security of the network. Based on the proposed  strategy, each sensor can directly exclude suspicious sensors from its neighboring set without detecting the information of all its neighbor sensors, which is more energy efficient than the existing works \cite{guan2017distributed,zhou2022security,yang2019distributed}. The practical experiment carried out on the three-phase electrical system also verified this conclusion. 
 }
	\item{To solve the NP-hard suspicious sensor selection problem, the original function is transformed into a submodular function (Lemma 3.1 and Theorem 3.1). On this basis, an attack detection scheduling algorithm based on the sequential submodular optimization theory is proposed (Algorithm 1). Moreover, in conjunction with \emph{expert problem}, the proposed algorithm can utilize historical information to guide the sensor selection task at the current moment, which improves the ability to deal with stealthy attacks.}
	\item{The proposed algorithm perfectly integrates the submodular theory with the problem of attack detection, which has not been attempted in the existing literature. 
 For the \emph{dynamic (static) attack strategy}, the proposed algorithm provides a higher theoretical lower bound of the average optimization rate than \cite{xu2022online} (Theorem 3.2 and Corollary 3.1).
 Finally, from the perspective of augmented estimation error, the proposed algorithm guarantees the security of the entire network under two general insecure conditions (Theorem 3.3). The effectiveness of the proposed algorithm is confirmed via numerical simulations.}
\end{itemize}

\emph{Notations}: 
Throughout this paper, $\mathbb{R} ^{n}$ and $\mathbb{R} ^{n\times n}$ represent the $n$-dimensional Euclidean space and $n\times n$ real matrices, respectively. $\mathbb{E}(\cdot )$ and $Pr(\cdot )$ refer to
the mathematical expectation and the probability, respectively.
For a matrix $A$, $||A||$ and $||A||_1$ separately stand for the $l_2$-norm and $l_1$-norm, while $A^T$ denotes its transpose.
The symbol $\otimes $ denotes the Kronecker product. And for the set $\mathcal{N}$, the $|\mathcal{N}|$ denotes its cardinality. 
For $j\in\mathcal{N}$, denote $\left[ w_{j} \right] _{j\in \mathcal{N}}$ as a $|\mathcal{N}|$-dimensional column vector composed of the values of all elements $w_j$, that is, $\left[ w_{j} \right] _{j\in \mathcal{N}}=[w_{1},...,w_{j},...,w_{|\mathcal{N}|}]^T$. Similarly, for $i\in\mathcal{N}$ and $j\in\mathcal{M}$, denote $\left[ w_{ij} \right] _{i\in \mathcal{N},j\in \mathcal{M}}$ as a $|\mathcal{N}|\times|\mathcal{M}|$-dimensional square matrix composed of the values of all elements $w_{ij}$.
The $|\mathcal{N}|$-dimensional diagonal matrix with the $i$-th position being $1$ is defined as $\theta_{|\mathcal{N}|}^i$.
The function $\lfloor \cdot \rfloor$ returns a number rounded down to a given number of places. 
In the following, the sensor network is regarded as a large-scale network, and the research is carried out.

\section{Problem formulation}

\subsection{System Model}
Consider a linear discrete-time linear system below:
\begin{equation} \label{state_equation}
	x\left( k+1 \right) =Ax\left( k \right) +\omega \left( k \right) ,
\end{equation}
where $x\left( k \right)\in \mathbb{R} ^{n}$ and $\omega \left( k \right)\in \mathbb{R} ^{n}$ represent the state of the system and process noise, respectively. And $\omega \left( k \right)$ follows Gaussian distribution with zero-mean and covariance matrix $Q>0$, i.e., $\omega \left( k \right) \sim \mathcal{N} \left( 0,Q \right)$. 

Suppose there is a large-scale sensor network composed of a series of sensors to monitor the state $x\left( k \right)$. Consider the network to be an undirected graph $ \mathcal{G} = (\mathcal{N}, \mathcal{E}) $, where $\mathcal{N}$ and $\mathcal{E} \subseteq \mathcal{N} \times \mathcal{N} $ represent the set of sensors and edges, respectively. The neighboring set of sensor $i$ is denoted by $\mathcal{N}_i = \{j \in\mathcal{N} : (i, j) \in \mathcal{E}\}$. Therefore, we can get $\mathcal{N}=\mathcal{N}_1 \cup \mathcal{N}_2 \cup \ldots \cup \mathcal{N}_{|\mathcal{N}|}$. For sensor $i$, $i \in \{1,2,...,|\mathcal{N}|\}$, the measurement model is:
\begin{equation} \label{measurement_equation}
	y_{i}\left( k \right) =C_{i}x\left( k \right) +\nu_{i} \left( k \right) ,
\end{equation}
where $y_{i}\left( k \right)\in \mathbb{R} ^{m}$ and $\nu_{i} \left( k \right)\in \mathbb{R} ^{m}$ represent the measurement of  sensor $i$ and measurement noise, respectively. Both $A\in\mathbb{R} ^{n\times n}$ and $C_i\in\mathbb{R} ^{m\times n}$ are real matrices with proper dimensions. Similarly, $\nu_{i} \left( k \right)$ follows Gaussian distribution with zero-mean and covariance matrix $R_{i}>0$, i.e., $\nu_{i} \left( k \right) \sim \mathcal{N} \left( 0,R_{i} \right)$.  In this paper, both process noise $||\omega \left( k \right)||$ and measurement noise $||\nu_{i} \left( k \right)||$ are upper bounded by small positive scalars.
For the neighboring set $\mathcal{N}_i$ of sensor $i$, it is assumed that $( A, [C_{1}^{T},...,C_{|\mathcal{N}_i|}^{T}]^{T} ) $ is observable.

Then, the distributed estimator of sensor $i$ is given by:
\begin{multline}
	\hat{x}_i\left( k+1 \right) =A\hat{x}_i\left( k \right) +K_i\left( k \right) \left( y_i\left( k \right) -C_i\hat{x}_i\left( k \right) \right) \\
	-\lambda A\sum_{j\in \mathcal{N} _i}{ \left( \hat{x}_i\left( k \right) -\hat{x}_{j}\left( k \right) \right)},
\end{multline}
where $\hat{x}_i(k)$ is the estimate of the state $x(k)$ of sensor $i$ with $\hat{x}_i(0) = \mathbb{E} \{x(0)\}$, $\hat{x}_j(k)$ is the estimate received from sensor $j$, $K_i(k)$ is the gain matrix, and $\lambda$ is the consensus parameter within $(0, \min(1/|\mathcal{N}_i| ))$, $\forall i\in \mathcal{N} $.

\subsection{Attack Model}
The attacker considered in this paper is an intelligent attacker that has access to all historical transmission data and has the ability to perform FDIAs on some of the communication links. At moment $k$, suppose that the estimate $\hat{x}_j(k)$ of sensor $j$ is injected with false data $z_{ij}(k)$ during its transmission to $i$. Then, the impaired estimate received by sensor $i$ is 
\begin{equation}\label{attack_signal}
	\hat{x}_{ij}^{a}\left( k \right) =\hat{x}_j\left( k \right) +z_{ij}\left( k \right) ,
\end{equation}
where $z_{ij}(k)$ is the false data injected by the attacker. 
For the attacker's attack strategy, we have the following assumption:

 \noindent {\bf Assumption 2.1 \cite{yang2021secure,lu2023polynomial}} At moment $k$, 
 the maximum number of attacked neighboring sensors around sensor $i$ is no more than half the number of neighboring sensors, that is, $q_i\le \lfloor \mathcal{N}_i/2 \rfloor$, where $q_i$ is defined as the number of the attacked sensors among the neighbor sensors of sensor $i$.

\noindent {\bf Remark 2.1} Assumption 2.1 has been verified in the literature \cite{yang2021secure,lu2023polynomial}. Indeed, this general assumption is to ensure that each sensor can achieve decision consistency \cite{an2017secure}. If the number of attacked sensors is less than or equal to half of the number of neighboring sensors, each sensor can achieve decision consistency and correctly estimate the system state through appropriate algorithms and protocols. However, when the number of attacked sensors exceeds half of the number of neighbor sensors, it is theoretically impossible for each sensor to correctly estimate the system state.

 \noindent {\bf Definition 2.1 (\emph{Dynamic attack strategy})} Define the attacked sensor sets at two adjacent moments as $\mathcal{A}_{k}$ and $\mathcal{A}_{k-1}$. Then, the difference of the attacked sensor group at adjacent moments can be calculated as  $\varDelta _k=( \mathcal{A}_{k}\backslash \mathcal{A}_{k-1} ) \cup ( \mathcal{A}_{k-1}\backslash \mathcal{A}_{k} ) $. Therefore, for the entire time step $T$, the number of changes in attack strategies is $\varDelta _T=\sum_{k=1}^{T-1}{\varDelta _k}$.

 \noindent {\bf Remark 2.2 (\emph{Static attack strategy})} The difference between the \emph{static attack strategy} and the \emph{dynamic attack strategy} is that the attacked sensor set under the \emph{static attack strategy} is fixed, while the attacked sensor set under the \emph{dynamic attack strategy} changes dynamically. Therefore, the \emph{static attack strategy} is essentially a special case of the \emph{dynamic attack strategy} in Definition 2.1, that is, $\varDelta _T=0$ is established.

 \noindent {\bf Remark 2.3} 
With the development of cryptography, end-to-end security approaches represented by pre-shared keys or certificate-based security have been deployed in some sensor networks.
 However, the existing literature and technical report indicate that keys and certificates can be compromised \cite{kwon2022certificate,mouha2021review}, which means that even in this case malicious attackers can launch FDIAs to tamper with the information transmitted between sensors. Therefore, the FDIAs can still occur in the real world.

To verify the authenticity of the received estimates, it is assumed that sensor $i$ is equipped with a detector \cite{zhou2022security},
\begin{equation}\label{detector}
	D_{ij}\left( k \right) =\left\| \hat{x}_i\left( k \right)- \hat{x}_{ij}^{a}(k)\right\| \underset{\mathcal{H} _1}{\overset{\mathcal{H} _0}{\lessgtr}}\upsilon _i\xi _i\left( k \right) ,
\end{equation}
where $\upsilon _i$ is a positive number, and at each moment $k$,  $\xi_{i}(k)$ is randomly generated by the detector, which obeys a random variable with exponential distribution with parameter $1$, i.e., $\xi _i(k)\sim E\left( 1 \right) $.
And the hypothesis $\mathcal{H}_0$ indicates  that the estimate of sensor $j$ received by sensor $i$ has not been tampered with, while the hypothesis $\mathcal{H}_1$ indicates the opposite. Define $\gamma _{ij}\left( k \right)$ as a binary variable representing the judgement of the detector at each moment $k$, that is
\begin{equation}
	\gamma _{ij}\left( k \right) =\left\{ \begin{array}{c}
		1, ~~\text{if}\,\,D_{ij}\left( k \right) \le \upsilon _i\xi _i\left( k \right) ,\\
		\!\!\!\!\!\!\!\!\!\!	\!\!\!\!\!	\!\!\!\!\!	\!\!\!\!\!	\!\!\!			0, ~~\text{otherwise}.\\
	\end{array} \right.
\end{equation}

Thus, the distributed estimator with the detector in sensor $i$ is given by:
\begin{multline}\label{distri_estimator_with_detector}
	\hat{x}_i\left( k+1 \right) =A\hat{x}_i\left( k \right) +K_i\left( k \right) \left( y_i\left( k \right) -C_i\hat{x}_i\left( k \right) \right) 
	\\
	-\lambda A\sum_{j\in \mathcal{N} _i}{\gamma _{ij}\left( k \right) \left( \hat{x}_i\left( k \right) -\hat{x}_{ij}^a\left( k \right) \right)}.
\end{multline}
To avoid ambiguity, the following $\hat{x}_i(k)$ are all calculated by the above equation if not otherwise specified. 

\subsection{Submodular Function Property}
For a sensor set $\mathcal{N}$, the function $f$ on it assigns a real value to each subset of $\mathcal{N}$, i.e., $f : 2^{\mathcal{N}} \rightarrow \mathbb{R}$.

\noindent {\bf Definition 2.1. \cite{matsuoka2021tracking}} 
For all $\mathcal{A}\subseteq \mathcal{B}\subseteq \mathcal{N}$, if the relationship $f(\mathcal{A}) \leq f(\mathcal{B})$ holds,  the function $f$ is monotone non-decreasing.

\noindent {\bf Definition 2.2. \cite{matsuoka2021tracking}} For every $\mathcal{A}\subseteq \mathcal{B}\subseteq \mathcal{N}$ and $j\in \mathcal{N}\backslash \mathcal{B}$, if the relationship 
\begin{equation}\label{sub_property}
    f\left( \mathcal{A}\cup \left\{ j \right\} \right) -f\left( \mathcal{A} \right) \ge f\left( \mathcal{B}\cup \left\{ i \right\} \right) -f\left( \mathcal{B} \right)
\end{equation}
holds,  the function $f$ is submodular.

Moreover, the function $f$ has the property of diminishing marginal returns, which means that the contribution of newly selected sensor to $f(\mathcal{A})$  decreases as more sensors are selected into the set $\mathcal{A}$.

\subsection{Problem of Interest}
For large-scale networks, the energy and computing power of each edge node are limited. However, current research on attack detection generally detects the information of all neighboring sensor, which brings a lot of energy consumption. Therefore, this paper considers reducing the energy and computing power consumption of attack detection by directly searching for the set of suspicious sensors. However, it is NP-hard to directly find the optimal set of suspicious sensors.
The following problems need to be investigated: how to improve the efficiency of finding the set of suspicious sensors, and whether each sensor only excludes some suspicious sensors from its neighboring sensors can ensure the security of the entire network.

\section{Main Results}
In this section, a novel security defense strategy from the perspective of attack detection scheduling is investigated. 
First, the problem of finding the set of suspicious sensors is proven to be NP-hard. To solve this problem, the original function is transformed into a submodular function. Then, an attack detection scheduling algorithm based on the sequential submodular optimization theory is proposed, and the theoretical lower bound of the its performance is also proved. Finally, it is proved that the proposed algorithm can guarantee the security of the entire distributed network.

\subsection{Problem Conversion}

Assuming that sensor $i$ is not equipped with a detector (no coefficient $\gamma _{ij}( k )$ in (\ref{distri_estimator_with_detector})), it can be seen from equation (\ref{distri_estimator_with_detector}) that $\sum_{j\in \mathcal{N} _i}{ ( \hat{x}_i\left( k \right) -\hat{x}_{ij}^a\left( k \right) )}$ has the greatest influence on the estimate of sensor $i$ at moment $k$.

Based on the above analysis, for the $j$-th sensor around sensor $i$, $j\in \mathcal{N}_i$, we use a new parameter $\mu_{ij}(k)$ to indicate whether sensor $j$ is in the set of suspicious sensors $\mathcal{A}_{i,k}$ at moment $k$. Thus, the original function can be obtained
    \begin{multline}\label{origin_func} 
  \sum_{j\in \mathcal{N} _i}{\mu _{ij}\left( k \right)}\left\| \hat{x}_i\left( k \right) -\hat{x}_{ij}^{a}(k) \right\| =
		\\
		\sum_{j\in \mathcal{A}_{i,k}}{\mu _{ij}\left( k \right) \left\| \hat{x}_i\left( k \right) -\hat{x}_{ij}^{a}(k) \right\| _{\mu _{ij}\left( k \right) =1}}+
		\\
		\sum_{j\in \left( \mathcal{N} _i\backslash \mathcal{A}_{i,k} \right)}{\mu _{ij}\left( k \right) \left\| \hat{x}_i\left( k \right) -\hat{x}_{ij}^{a}(k) \right\| _{\mu _{ij}\left( k \right) =0}}
	\end{multline}
where   $\hat{x}_{ij}^{a}(k)$ and $\hat{x}_i\left( k \right)$ are obtained from  (\ref{attack_signal}) and (\ref{distri_estimator_with_detector}), respectively.

In the following, Lemma 3.1  shows that the problem of finding the set of suspicious sensors is NP-hard and transforms (\ref{origin_func}) into a solvable form.

\noindent {\bf Lemma 3.1.} 
For sensor $i$, the original function (\ref{origin_func}) can be  transformed into the following solvable form
	\begin{equation} \label{f_A}
	f_k\left( \mathcal{A}_{i,k} \right) =\left\| \varLambda_{i,k} \cdot [\mu_{ij}(k)]_{j\in\mathcal{N}_i} \right\|,
\end{equation}
where   $[\mu_{ij}(k)]_{j\in\mathcal{N}_i}$ represents the augmented matrix of $\mu_{ij}(k)$, and for $j\in \mathcal{A}_{i,k}$, $\mu_{ij}(k)=1$, for $j\in \mathcal{N}_i\backslash\mathcal{A}_{i,k}$, $\mu_{ij}(k)=0$. The parameter $\varLambda_i$ is the  augmented error matrix, which is described in detail in the proof.
\begin{proof}
	 For sensor $j\in\mathcal{N}_i$, it can be seen from the original function (\ref{origin_func}) that whether the sensor $j$ belongs to the set of suspicious sensors $\mathcal{A}_{i,k}$ is a binary hypothesis.
And we find that the selected set $\mathcal{A}_{i,k}$ of the $0$-$1$ knapsack problem is optimal if and only if the set $\mathcal{A}_{i,k}$ can maximize the original function (\ref{origin_func}). However, finding the optimal solution to the $0$-$1$ knapsack problem is generally NP-hard.
 Therefore, we consider transforming (\ref{origin_func}) into a solvable form.

The augmented error matrix $\varLambda_{i,k}$ is a diagonal matrix that summarizes the errors $||\hat{x}_i( k )- \hat{x}_{ij}^{a}(k)||$ of all neighboring sensors around sensor $i$, which can be calculated as
\begin{equation}\label{residual_matrix}
	\varLambda_{i,k} =\sum^{|\mathcal{N}_i|}_{j=1}{\left(\theta_{|\mathcal{N}_i|}^j\otimes||\hat{x}_i\left( k \right) -\hat{x}_{ij}^{a}\left( k \right)||\right)},
\end{equation}

	 Then, the original function (\ref{origin_func}) can be transformed into
  \begin{multline}
      f_k\left( \mathcal{A}_{i,k} \right) =\left\| \varLambda_{i,k} \cdot [\mu_{ij}(k)]_{j\in\mathcal{N}_i} \right\|\\
	=\| \sum^{|\mathcal{N}_i|}_{j=1}{\left(\theta_{|\mathcal{N}_i|}^j\otimes||\hat{x}_i\left( k \right) -\hat{x}_{ij}^{a}\left( k \right)||\right)} \cdot [\mu_{ij}(k)]_{j\in\mathcal{N}_i} \|.
  \end{multline}	
 where $[\mu_{ij}(k)]_{j\in\mathcal{N}_i}$ represents the augmented matrix of $\mu_{ij}(k)$,
 and for $j\in \mathcal{A}_{i,k}$, $\mu_{ij}(k)=1$, for $j\in \mathcal{N}_i\backslash\mathcal{A}_{i,k}$, $\mu_{ij}(k)=0$.

This completes the proof.
\end{proof}

From the above Lemma 3.1, it can be seen that the problem of maximize the original function (\ref{origin_func}) can be transformed into finding no more than $q_i$ sensors  in the set $\mathcal{N}_i$ that have the greatest influence on (\ref{f_A}) at moment $k$, which can be described as {\bf{Problem 1}} below:
\begin{flalign} \label{problem1}
	&\max_{\mathcal{A}_{i,k} \subseteq \mathcal{N}_i} f_k\left( \mathcal{A}_{i,k} \right) ~~\text{s.t.} \left| \mathcal{A}_{i,k} \right|\leq q_{i},&
\end{flalign}
where $q_{i}$ is defined in Assumption 2.1. It should be noted that (\ref{origin_func}) and (\ref{f_A}) are not essentially equivalent, but they both aim to select suspicious sensors.

The following Theorem 3.1 proves that the \emph{objective function} (\ref{f_A}) is a submodular function, which lays the foundation for the subsequent algorithm.

\noindent {\bf Theorem 3.1.} 
For sensor $i$, the problem of finding the set of suspicious sensors $\mathcal{A}_{i,k}$ satisfies the properties of submodular function, that is, 
 the \emph{objective function} $f_k\left( \mathcal{A}_{i,k} \right) $ in  (\ref{f_A}) is monotonically non-decreasing and submodular.
 \begin{proof}
Consider two sets $\mathcal{A}_{i,k} $ and $\mathcal{B}_{i,k} $ belonging to $\mathcal{N}_i $, and all elements in $\mathcal{A}_{i,k} $ are contained in $\mathcal{B}_{i,k} $, that is, $\mathcal{A}_{i,k} \subseteq \mathcal{B}_{i,k} $ is established.
 
     First, for the monotonically non-decreasing property of $f_k(\mathcal{A}_{i,k})$ in Definition 2.1, 
	since the augmented error matrix $\varLambda_{i,k} $ is a diagonal matrix, the value of $f_k(\mathcal{A}_{i,k})$ can be equivalently obtained by summing the squares of the error terms corresponding to the elements contained in the set $\mathcal{A}_{i,k}$ and then extracting the square root. 
	Therefore, for $\mathcal{A}_{i,k} \subseteq \mathcal{B}_{i,k} $, $f_k\left( \mathcal{B}_{i,k} \right) -f_k\left( \mathcal{A}_{i,k} \right)\ge 0$ always holds, that is, $f_k(\mathcal{A}_{i,k})$ is monotonically non-decreasing.

 Then, to prove the submodular property of the \emph{objective function} $f_k(\cdot)$, it is necessary to prove that following inequality in Definition 2.2
\begin{equation}\label{prove_sub}
     f_k\left( \mathcal{A}_{i,k} \cup \left\{ j \right\} \right) -f_k\left( \mathcal{A}_{i,k} \right)\ge f_k\left( \mathcal{B}_{i,k} \cup \left\{ j \right\} \right) -f_k\left( \mathcal{B}_{i,k}\right)
 \end{equation} holds. 
We multiply both sides of (\ref{prove_sub}) by $(f_k\left( \mathcal{A}_{i,k} \cup \left\{ j \right\} \right) +f_k\left( \mathcal{A}_{i,k} \right)) (f_k\left( \mathcal{B}_{i,k} \cup \left\{ j \right\} \right) +f_k\left( \mathcal{B}_{i,k}\right))$. 

According to the relationship that $(f_k\left( \mathcal{A}_{i,k} \cup \left\{ j \right\} \right) -f_k\left( \mathcal{A}_{i,k}\right))(f_k{(\mathcal{A}_{i,k} \cup \left\{ j \right\} )} +f_k\left( \mathcal{A}_{i,k}\right))=f(\{j\})^2$, the proof of (\ref{prove_sub}) can be transformed into proving the inequality 
\begin{equation}
    f_k\left( \mathcal{B}_{i,k} \cup \left\{ j \right\} \right) +f_k\left( \mathcal{B}_{i,k} \right))\ge  f_k\left( \mathcal{A}_{i,k} \cup \left\{ j \right\} \right) +f_k\left( \mathcal{A}_{i,k} \right)
\end{equation}
holds, which is established because of the monotonically non-increasing property of $f_k(\mathcal{A}_{i,k})$.
	Thus, for $\mathcal{A}_{i,k} \subseteq \mathcal{B}_{i,k} $, inequality (\ref{prove_sub}) holds, that is, $f_k(\mathcal{A}_{i,k})$ is submodular.

 Based on the above proof, the \emph{objective function} $f_k\left( \mathcal{A}_{i,k} \right) $ in  (\ref{f_A}) is monotonically non-decreasing and submodular.
 This completes the proof.
 \end{proof}

\noindent {\bf Example 3.1.} Assume that there are $4$ neighbor sensors around sensor $i$, and their estimation errors are $1$, $2$, $3$, and $4$. Suppose there are two sets $\mathcal{A}_i$ and $\mathcal{B}_i$, which contain $\{4\}$ and $\{3, 4\} $ respectively. Then, the values of the \emph{objective function} (\ref{f_A}) are $\sqrt{4^2}$ and $\sqrt{3^2+4^2}$ respectively. Add element $\{2\}$ to the sets $\mathcal{A}_i$ and $\mathcal{B}_i$, then the gain changes brought about by adding the element are $\sqrt{2^2+4^2}-\sqrt{4^2}=0.47$ and $\sqrt{2^2+3^2+4^2}-\sqrt {3^2+4^2}=0.38$. Obviously, $0.38<0.47$, which is consistent with the conclusion of Theorem 3.1.

\subsection{The Design of Attack Detection Scheduling Algorithm Based on the Sequential Submodular Optimization Theory}

The aforementioned Theorem 3.1 proves that the \emph{objective function} (\ref{f_A}) is a submodular function, so the submodular optimization theory in \cite{matsuoka2021tracking} can be used to select the  set of suspicious sensors.
In recent years, the sequential submodular optimization algorithm has extended the submodular optimization algorithm from a single moment to the entire time step, and has been applied in the fields of robot scheduling\cite{xu2022resource}, target tracking\cite{tzoumas2020robust}, and so on. And the large-scale network security problem investigated in this paper also require historical information to guide the selection of suspicious sensors at the current moment. Therefore, this paper considers to propose an attack detection scheduling algorithm based on the sequential submodular optimization theory.

As shown in Algorithm \ref{alg:Seq_Sub_Max_new} below, for sensor $i$, at the $l$-th selection at each moment $k$, an empty set is created and the gains of all remaining neighbouring sensors are calculated. Drawing on the \emph{expert problem} in \cite{herbster1998tracking}, the coefficient $\beta$ is used to weigh the information of the past moment and the current moment. Then, the $l$-th selection is selected into the candidate set according to the vector of distribution proportions $p_k^{(l)}$.
The algorithm ends when there are $q_i$ sensors in the candidate set $\mathcal{A}_{i,k}^{(l)}$.

\begin{algorithm}[htb]  
	\caption{The Attack Detection Scheduling Algorithm Based on the Sequential Submodular Optimization Theory}
	\label{alg:Seq_Sub_Max_new}  
	\begin{algorithmic}[1]  
		\Require   
  The neighbouring set $\mathcal{N}_i$ of sensor $i$, the maximum number of attacked neighboring sensors $q_i$, coefficient $\beta$, historical information value $W_{kj}$, $j\in\mathcal{N}_i$.   
		\Ensure  
		The set of suspicious sensors $\mathcal{A}_{i,k}$ at moment $k$, $k=1,2,...,T$.

		\State Initialize weight vector 
  $\omega_{k}^{(l)}=[\omega_{kj}^{(l)}]_{j\in\mathcal{N}_{i}}$
  by $\omega _{kj}=1$, $\mathcal{A}_{i,k}^{(0)}=\emptyset$ for $k=1,...,T$, $l=1$.
    \While{$l<q_i$}
    \State Set $\mathcal{S}_{k}^{(l)}=\mathcal{A}_{i,k}^{(l)}=\emptyset$.
        \For {all $j\in \mathcal{N}_i\backslash\mathcal{A}_{i,k}^{(l-1)}$}
		\State Calculate $G_{kj}^{(l)} = f_k( \mathcal{A}_{i,k}^{(l-1)} )-f_k( \mathcal{A}_{i,k}^{(l-1)}\cup \left\{ j \right\} ) $.
    \State Calculate $v_{kj}^{(l)}=w_{k,j}^{(l-1)}e^{ -G_{kj}^{(l)}}$.
  \If{$\beta=0$}
  	\State Update $w_{k}^{(l)}$ by $w_{kj}^{(l)}= v_{kj}^{(l)}$.
  \Else
		\State Update $w_{k}^{(l)}$ by $w_{kj}=\beta \frac{W_{kj}}{k-1}+\left( 1-\beta \right) v_{kj}^{(l)}$.
  \EndIf
  \EndFor		
  \State Calculate $p_{k}^{(l)}=w_{k}^{(l)}/\| w_{k}^{(l)}\|_1$.
\State Draw an item $j^{(l)}_{select}$ from the distribution $p_{k}^{(l)}$.
\State Get $\mathcal{A}_{i,k}^{(l)}=\mathcal{A}_{i,k}^{(l-1)}\cup \{j_{,select}^{(l)}\}$.
\State $l=l+1$.
\EndWhile

\State \Return $\mathcal{A}_{i,k}=\mathcal{A}_{i,k}^{(l)}$.

	\end{algorithmic}  
\end{algorithm}

The summary of the variable notations in Algorithm \ref{alg:Seq_Sub_Max_new} is shown in TABLE \ref{tab1}.
To formally describe the algorithm, we make the following notes:
\begin{itemize}
\item{The historical information value $W_{kj}$  is calculated by $W_{kj}=\sum^{k-1}_{t=1}{e^{-\frac{1}{t}}v_{tj}^{(0)}}$, which is only related to $v_{tj}^{(0)}$  at moment $t=1,...,k-1$. And  the reason for setting the coefficient of $v_{tj}^{(0)}$ to $\exp(-1/t)$ is that the most recently attacked sensor is considered to have greater weight.}
\item{In steps $7-11$, the value of the coefficient $\beta$ determines whether to draw on \emph{expert problem} to guide the $l$-th selection at moment $k$. In step $10$, we draw on the \emph{expert problem} in \cite{herbster1998tracking}.}
\item{In step $14$, we select $j_{select}^{(l)}$ according to the vector of distribution proportions $p_k^{(l)}$. To improve the accuracy of selection, we can complete the sorting before each selection and then directly find the maximum value, but the computational complexity will also increase, which will be discussed later.}
\end{itemize}

\begin{table}[!h]
	\caption{Summary of variable notations in algorithm \ref{alg:Seq_Sub_Max_new}}
	\centering
	\begin{tabular}{|l||l|}
		\hline
		Variables & Meanings\\
		\hline
		$\omega_k^{(l)}$ & The weight vector $\omega_k^{(l)}$ to be updated     \\
     & at the $l$-th selection at moment $k$ \\
		\hline
		$\mathcal{N}_i\backslash\mathcal{A}_{i,k}^{(l-1)}$ & The set of remaining sensors after excluding the\\
		  &  selected sensors at the $l$-th selection at moment $k$
		    \\
		\hline
		$f_k(\cdot)$ & The objective function at moment $k$\\
		\hline
		$ \mathcal{A}_{i,k} $ & The candidate set at moment $k$\\
		\hline
		$G_{k}^{(l)}$ & The gain vector at the $l$-th selection at moment $k$, \\
  & and $G_{k}^{(l)}=[G_{kj}^{(l)}]_{j\in\mathcal{N}_i\backslash\mathcal{A}_{i,k}^{(l-1)}}$\\
		\hline
		$\beta$ & The coefficient to weigh the past information and the \\
		  &   present moment information  \\
		\hline
		 $p_k^{(l)}$ & $p_k^{(l)}$ is a vector of distribution proportions \\
   & at the $l$-th selection at moment $k$,\\
 &   which can be obtained by normalizing the  vector $\omega_k^{(l)}$ \\
		\hline
		$j_{select}^{(l)}$ & The selected sensor at the $l$-th selection\\
		\hline
		$v_{kj}^{(l)}, W_{kj} $ & The intermediate variables generated by the \emph{expert} \\
		  &  \emph{problem} when updating the weight vector 	$\omega_k^{(l)}$\\
		\hline
	\end{tabular}
    \label{tab1}
\end{table}

\noindent {\bf Proposition 3.1.} (Complexity) For sensor $i$, at the $l$-th selection at moment $k$, the proposed algorithm requires $ |\mathcal{N} _i\backslash \mathcal{A}_{i,k}^{(l-1)}| $ times evaluation operations to  calculate gain and weight vectors, $1$ time of normalization operation and selection operation. This means that the  complexity of the proposed algorithm is approximately $\mathcal{O}(q_i|\mathcal{N}_i|)$.

In the following, we theoretically illustrate the performance of the Algorithm \ref{alg:Seq_Sub_Max_new} by Theorem 3.2.  Before the proof of Theorem 3.2, the following Lemma 3.2 is introduced.

\noindent {\bf Lemma 3.2.\cite{matsuoka2021tracking}} 
For  $l\in\{0,1,...,q_i\}$, define $\delta_l$ as $\delta_l= \sum_{k=1}^T({f_k( \mathcal{A}_{i,k}^{*} )}-{f_k( \mathcal{A}_{i,k}^{(l)} )})$
,  where $ \mathcal{A}_{i,k}^{*} $ and $\mathcal{A}_{i,k}^{(l)}$ respectively represent the optimal candidate set of sensor $i$ and the candidate set after the $l$-th selection at moment $k$. Denote $B_{i}^{\left( l \right)}$ as $B_{i}^{\left( l \right)}=\sum_{j=1}^{q_i}\sum_{k=1}^T{( G_{k}^{\left( l \right)}p_{k}^{\left( l \right)}-G_{kj_{kl}^{*}}^{\left( l \right)} )}$, where $G_k^{(l)}=[G_{kj}^{(l)}]_{j\in\mathcal{N}_i\backslash\mathcal{A}_{i,k}^{(l-1)}}$ and $G_{kj_{kl}^{*}}^{\left( l \right)}$ represents the optimal gain when selecting the optimal sensor $j_{kl}^{*}$, for the $l$-th sensor selection at moment $k$.
Then, the relationship between $B_i^{(l)}$ and $\delta_l$, $l\in\{0,1,...,q_i\}$, is
\begin{equation}\label{Lemma32}
	\delta_{q_i} -(1-\frac{1}{q_i})^{q_i}\delta_0 \le \frac{1}{q_i}\sum_{l=1}^{q_i}(1-\frac{1}{q_i})^{q_i-l} B_i^{(l)}.
\end{equation}

\begin{proof}
For an arbitrary fixed $l\in \{0,1,...,q_i-1\}$, we have
\begin{eqnarray}
	\delta_l&=&\sum_{k=1}^T({f_k( \mathcal{A}_{i,k}^{*} )}-{f_k( \mathcal{A}_{i,k}^{(l)})})\nonumber\\
	&\le& \sum_{k=1}^T \sum_{j\in \mathcal{A}_{i,k}^*} ( f_k(\mathcal{A}_{i,k}^{(l)}\cup \{j_{kl}^*\})-f_k(\mathcal{A}_{i,k}^{(l)}) ) \nonumber\\
	&=&\sum_{k=1}^T (-\sum_{j \in \mathcal{A}_{i,k}^*}G_{kj}^{(l+1)})
	=-q_i\sum_{k=1}^T G_{kj_{k,l+1}}^{(l+1)}+B_i^{(l+1)}\nonumber\\
	&=&q_i(\delta_l-\delta_{l+1})+B_i^{(l+1)},
\end{eqnarray}
where the inequality holds because of the submodularity of $f_k$,  the second equality follows from the definition of $G_{kj}^{(l+1)}$, the third equality follows from the definition of $B_i^{(l+1)}$, and the fourth equality follows from the definition of $\delta_l$.

Thus, we have
\begin{equation}
	\delta_{l+1} \le (1-\frac{1}{q_i})\delta_l+\frac{1}{q_i}B_i^{(l+1)},
\end{equation}
holds for each $l \in \{0,1,...,q_i\}$, and hence, we have
\begin{equation}
	\delta_{l+1} \le (1-\frac{1}{q_i})^{l+1}\delta_0+\frac{1}{q_i}\sum_{j=1}^{l+1}(1-\frac{1}{q_i})^{l+1-j} B_i^{(j)}.
\end{equation}

Therefore, $\delta_{q_i} -(1-\frac{1}{q_i})^{q_i}\delta_0 \le \frac{1}{q_i}\sum_{l=1}^{q_i}(1-\frac{1}{q_i})^{q_i-l} B_i^{(l)}$ is established.
 This completes the proof.
\end{proof}

The evaluation metrics are introduced in Definition 3.3.

\noindent {\bf Definition 3.3.} At moment $k$, the optimization rate is defined as the ratio of the \emph{objective function} (\ref{f_A}) value of the set of suspicious sensors selected by Algorithm \ref{alg:Seq_Sub_Max_new} to the optimal \emph{objective function} (\ref{f_A}) value, which is defined as $f_k\left( \mathcal{A} _{i,k} \right) /f_k( \mathcal{A} _{i,k}^{*} )$,  where $\mathcal{A} _{i,k} $ and $\mathcal{A} _{i,k}^{*}$ represent the set of suspicious sensors selected by Algorithm \ref{alg:Seq_Sub_Max_new} and the optimal  set of suspicious sensors for sensor $i$ at moment $k$, respectively. And for the entire time period $T$, the average optimization rate can be defined as $\frac{1}{T}\sum_{k=1}^T{( f_k( \mathcal{A} _{i,k} ) /f_k( \mathcal{A} _{i,k}^{*} ) )}$.
Therefore, we only need to prove that the proposed algorithm can guarantee the theoretical lower bound on the average optimization rate to show the performance of the proposed algorithm over the entire time period.

In the following, we illustrate the theoretical lower bound on the average optimization rate of the algorithm \ref{alg:Seq_Sub_Max_new}. 

\noindent {\bf Theorem 3.2.} 
For the dynamic attack strategy in Definition 2.1, the theoretical lower bound on the average optimization rate of the proposed attack detection scheduling algorithm is $1-1/e$, and the error expectation is bounded by 
	\begin{multline}
		\mathbb{E} [ ( 1-\frac{1}{e} ) \sum_{k=1}^T{f_k\left( \mathcal{A}_{i,k}^{*} \right)}-\sum_{k=1}^T{f_k( \mathcal{A}_{i,k} )} ]\\
  \le \tilde{\mathcal{O}}( \sqrt{q_iT\left( 2q_i+\varDelta _T \right)} ),
	\end{multline}
where $f_k( \mathcal{A}_{i,k}^{*})$ and $f_k\left( \mathcal{A}_{i,k} \right)$ denote the \emph{objective function} value of the best candidate set of sensor $i$ and the \emph{objective function} value of the candidate set selected by the proposed algorithm at moment $k$, respectively. 
\begin{proof}
To demonstrate that the theoretical lower bound on the optimization rate over the entire time step is
		 $1-1/e $,    
   it is necessary to prove that the following equation  \begin{equation}\label{difference_between_real_and_predict}
		\mathbb{E} [ ( 1-\frac{1}{e} ) \sum_{k=1}^T{f_k\left( \mathcal{A}_{i,k}^{*} \right)}-\sum_{k=1}^T{f_k\left( \mathcal{A}_{i,k} \right)} ] 
	\end{equation}
	is bounded, where $f_k\left( \mathcal{A}_{i,k} \right)$ denotes the value of the submodular function of the candidate set selected by the proposed algorithm at moment $k$.

 Then, according to \cite{matsuoka2021tracking}, we have
	\begin{multline}\label{Theorem32_25}
		( 1-\frac{1}{e} ) \sum_{k=1}^T{f_k\left( \mathcal{A}_{i,k}^{*} \right)}-\sum_{k=1}^T{f_k\left( \mathcal{A}_{i,k} \right)}\\
		\le(1-(1-\frac{1}{q_i})^{q_i})\sum_{k=1}^T{f_k\left( \mathcal{A}_{i,k}^{*} \right)}-\sum_{k=1}^T{f_k\left( \mathcal{A}_{i,k} \right)}\\
		\le \sum_{k=1}^T{f_k\left( \mathcal{A}_{i,k}^{*} \right)}-\sum_{k=1}^T{f_k\left( \mathcal{A}_{i,k} \right)}-(1-(1-\frac{1}{q_i})^{q_i})\sum_{k=1}^T{f_k( \mathcal{A}_{i,k}^{*} )}\\
		+(1-(1-\frac{1}{q_i})^{q_i})\sum_{k=1}^T{f_k( \mathcal{A}_{i,k}^{(0)} )}\le \delta_{q_i}-(1-\frac{1}{q_i})^{q_i}\delta_0,
	\end{multline}
where the first inequality follows from $(1-1/k)^{k}\le 1/e$ when $k$ approaches infinity, and the second inequality follows from the properties of submodular function $f_k$.

Combining (\ref{Lemma32}) and (\ref{Theorem32_25}), we can obtain
\begin{multline}
    ( 1-\frac{1}{e} ) \sum_{k=1}^T{f_k\left( \mathcal{A}_{i,k}^{*} \right)}-\sum_{k=1}^T{f_k\left( \mathcal{A}_{i,k} \right)}\\
    \le \frac{1}{q_i}\sum_{j=1}^{q_i}(1-\frac{1}{q_i})^{q_i-j} B_i^{(j)}.
\end{multline}

Thus, to show that (\ref{difference_between_real_and_predict}) is bounded, we only need to prove that $\mathbb{E} [ B_{i}^{\left(l \right)}]$ is bounded, $l=1,...,q_i$, that is
 	\begin{eqnarray}
		&&\mathbb{E} [ B_{i}^{\left( l \right)} ] =\sum_{j=1}^{q_i}{\mathbb{E} [ \sum_{k=1}^T{( G_{k}^{\left( l \right)}p_{k}^{\left( l \right)}-G_{kj_{kl}^{*}}^{\left( l \right)} )} ]}
		\nonumber\\
  &\le&2 \sum_{j=1}^{q_i}{\sqrt{T\left( 2\left(\varDelta _k +1 \right) \log \left(|\mathcal{N}_i|T \right) +\log \left( T \right) \right)}}
		\nonumber\\
		&\le&2\sqrt{q_iT\sum_{j=1}^{q_i}{\left( 2\left(\varDelta _k+1 \right) \log \left( |\mathcal{N}_i|T \right) +\log \left(  T \right) \right)}}
		\nonumber\\
		&\le&2\sqrt{q_iT\left( 2\left( \varDelta _T+q_i\log \left( |\mathcal{N}_i|T \right) \right) +q_i\log \left(  T \right) \right)},
	\end{eqnarray}
where  the first inequality comes from Corollary 1 in \cite{herbster1998tracking}, the second inequality comes from the Cauchy-Schwartz inequality, and the last inequality comes from $\varDelta _T=\sum_{k=1}^{T-1}{\varDelta _k}$ in Assumption 2.2. The above proof is similar to the literature \cite{matsuoka2021tracking}, except that the difference in parameter definitions leads to differences in the proof process.

Therefore, the equation (\ref{difference_between_real_and_predict}) is bounded
	\begin{multline}\label{difference_29}
		\mathbb{E} [ ( 1-\frac{1}{e} ) \sum_{k=1}^T{f_k\left( \mathcal{A}_{i,k}^{*} \right)}-\sum_{k=1}^T{f_k\left( \mathcal{A}_{i,k} \right)} ]\\
  \le \tilde{\mathcal{O}}( \sqrt{q_iT\left( 2q_i+\varDelta _T \right)} ),
	\end{multline}
where $\mathbb{E} \left[ \cdot \right] $ represents the expectation
taken with respect to the internal randomness, and $\tilde{\mathcal{O}}$ hides the log terms.
That is, the theoretical lower bound on the optimization rate of the proposed algorithm is obtained. 
This completes the proof.
\end{proof}

Theorem 3.2 proves that the theoretical lower bound of the average optimization rate of the proposed algorithm is $1-1/e$. The existing literature \cite{xu2022online} mainly focuses on the optimal action selection in the dynamic environment where the objective function $f_k(\cdot)$ is unknown, and the lower bound of its optimization rate is proved to be $1/2$. The proposed algorithm combines the sequential submodular optimization theory with the attack detection problem, that is, the submodular optimization theory is used to select suspicious sensors (at this time, the \emph{objective function} $f_k(\cdot)$ is known). In addition, the literature \cite{xu2022online} considers that any action has a gain, but the proposed algorithm considers the gain of the wrongly selected sensor to be $0$.

\noindent {\bf Corollary 3.1. } 
For the dynamic attack strategy in Definition 2.1, the theoretical lower bound on the average optimization rate of the proposed attack detection scheduling algorithm is $1-1/e$, and the error expectation is bounded by
	\begin{equation}
		\mathbb{E} [ ( 1-\frac{1}{e} ) \sum_{k=1}^T{f_k\left( \mathcal{A}_{i,k}^{*} \right)}-\sum_{k=1}^T{f_k\left( \mathcal{A}_{i,k} \right)} ] \le \tilde{\mathcal{O}}( q_i\sqrt{T} ),
	\end{equation}
where $f_k( \mathcal{A}_{i,k}^{*})$ and $f_k\left( \mathcal{A}_{i,k} \right)$ denote the \emph{objective function} value of the best candidate set of sensor $i$ and the \emph{objective function} value of the candidate set selected by the proposed algorithm at moment $k$, respectively. 

\begin{proof}
The static attack strategy is essentially a special case of the dynamic attack strategy in Definition 2.1, that is, $\varDelta _T=0$ is established. At this time, the proof of this corollary is similar to the proof of Theorem 3.2. Therefore, the proof is omitted.
\end{proof}

\subsection{Security Analysis}
To show that the proposed algorithm  can guarantee the security of the entire large-scale network, the augmented estimation error of the distributed system is proved to be bounded under two kinds of general insecurity conditions.

During the whole time steps, assuming there exists a virtual estimator that is not equipped with an attack detector, which means that malicious information will be used in the estimator update process, the distributed estimator can be written as 
\begin{multline}\label{distri_estimator_without_detector}
		\hat{x}_{i}^{\prime}\left( k+1 \right) =A\hat{x}_{i}^{\prime}\left( k \right) +K_i\left( k \right) \left( y_i\left( k \right) -C_i\hat{x}_{i}^{\prime}\left( k \right) \right) 
		\\
		-\lambda A\sum_{j\in \mathcal{N} _i}{( \hat{x}_{i}^{\prime}\left( k \right) -\hat{x}_{ij}^{\prime a}\left( k \right) )},
	\end{multline}
	where $\hat{x}_{ij}^{\prime a}\left( k \right)=\hat{x}_{j}^{\prime}\left( k \right)+z_{ij}(k)$.
	Define the state estimation difference between (\ref{distri_estimator_with_detector}) and (\ref{distri_estimator_without_detector}) as $\varDelta \hat{x}_i\left( k \right) =\hat{x}_{i}^{\prime}\left( k \right) -\hat{x}_i\left( k \right) $, we have	
	\begin{small}	\begin{multline}\label{difference_between_6_9}
			\varDelta \hat{x}_i\left( k+1 \right) =\left( A -K_i\left( k \right) C_i \right) \varDelta \hat{x}_i\left( k \right)\\
			+\lambda A\sum_{j\in \mathcal{N} _i}{\gamma _{ij}\left( k \right)}z_{ij}\left( k \right)
			-\lambda A\sum_{j\in \mathcal{N} _i}{\gamma _{ij}\left( k \right) \left( \varDelta \hat{x}_i\left( k \right) -\varDelta \hat{x}_j\left( k \right) \right)}.
		\end{multline}
	\end{small}
 
Inspired by the unsecurity conditions in \cite{zhou2022security,hu2018state}, the following Definitions 3.4 and 3.5 give two general unsecurity conditions.
 One with attack strategy that leads to $\lim_{\left\| z_{ij}\left( k \right) \right\| \rightarrow \infty} \left\| \varDelta \hat{x}_i\left( k+1 \right) \right\| \rightarrow \infty $, which is  known as general unstealthy attacks,  as shown in Definition 3.4. And another with strategy that leads to $\lim_{k\rightarrow \infty} \left\| \varDelta \hat{x}_i\left( k+1 \right) \right\| \rightarrow \infty $, which is known as general stealthy attacks,  as shown in Definition 3.5. 

\noindent {\bf Definition 3.4. } A system is insecure if there exists at least one attack strategy such that both of the following conditions are satisfied:
\begin{enumerate}
	\item For the state estimation difference $\varDelta \hat{x}_i\left( k+1 \right) $, we have
	\begin{equation}
		\lim_{\left\| z_{ij}\left( k \right) \right\| \rightarrow \infty} \left\| \varDelta \hat{x}_i\left( k+1 \right) \right\| \rightarrow \infty.
	\end{equation}
	\item The attack signal $z_{ij}$ injected by the attacker is unbounded.
\end{enumerate}

\noindent {\bf Definition 3.5.} A system is insecure if there exists at least one attack strategy such that both of the following conditions are satisfied:
\begin{enumerate}
	\item For the state estimation difference $\varDelta \hat{x}_i\left( k+1 \right) $, we have
\begin{equation}
		\lim_{k\rightarrow \infty} \left\| \varDelta \hat{x}_i\left( k+1 \right) \right\| \rightarrow \infty.
	\end{equation}
	\item The attack signal $z_{ij}$ injected by the attacker is bounded, that is
	\begin{equation}
		\left\| z_{ij}\left( k \right) \right\| \le \tilde{z}_i,
	\end{equation}
where $\tilde{z}_i$ is a small positive constant scalar.
\end{enumerate}

Then, the augmented estimation error describing the security of the entire network is defined.

\noindent {\bf Definition 3.6. \cite{zhou2022security}(Augmented estimation error)}	
 Combining the estimation errors in (\ref{difference_between_6_9}) of all $\mathcal{N}$ sensors, i.e., $\varDelta \hat{x}\left( k \right) \triangleq [ \begin{matrix}
		\varDelta \hat{x}_1^T\left( k \right),...,\varDelta \hat{x}_{|\mathcal{N}|}^T\left( k \right)\\
	\end{matrix} ]^T $. Then the augmented estimation error is given as follows:
	\begin{equation}\label{augmented state estimation error}
		\Delta \hat{x}(k+1)=F(k) \Delta \hat{x}(k)+\lambda(\Upsilon(k) \otimes A) Z(k),
	\end{equation}
	where $F(k) =\left(I_{\mathcal{N}}-\lambda \Gamma(k)\right) \otimes A-(\sum_{i=1}^{\mathcal{N}} \theta_{|\mathcal{N}|}^i \otimes K_i^{T}(k)) C$, $\Upsilon(k) =\operatorname{diag}\{\Upsilon_1(k), \ldots, \Upsilon_{|\mathcal{N}|}(k)\}$, $\Upsilon_i(k) =[\gamma_{i 1}^{\prime}(k), \ldots, \gamma_{i |\mathcal{N}|}^{\prime}(k)]$, $Z(k) =[z_1^T(k), \ldots, z_{|\mathcal{N}|}^T(k)]^T$, $z_i(k) =[z_{i 1}^T(k), \ldots, z_{i |\mathcal{N}|}^T(k)]^T, i \in\mathcal{N}$, $\Gamma(k) =\left[l_{i j}(k)\right]_{i,j\in\mathcal{N}}$, and
	\begin{equation}
		l_{i j}(k) = \begin{cases}-\gamma_{i j}(k),  \text { if }(i, j) \in \mathcal{E}, i \neq j, \\\nonumber
			-\sum_{j \in \mathcal{N}} \gamma_{i j}(k),  \text { if } i=j, \\\nonumber
			0,  \text { otherwise. }\end{cases}
	\end{equation}

\noindent {\bf Remark 3.2.} For sensor $i$, the estimation error $\varDelta \hat{x}_i\left( k \right) =\hat{x}_{i}^{\prime}\left( k \right) -\hat{x} _i\left( k \right) $ describes the difference in the state estimate of sensor $i$ with and without an attack detector at moment $k$.
 The augmented estimation error expresses the estimated difference of all sensors in the large-scale network in an augmented form. Therefore, the security of the entire network can be described by the augmented estimation error.

The following Theorem 3.3 gives the analysis of the effectiveness of the proposed algorithm under two insecurity conditions from the perspective of augmented estimation error.

\noindent {\bf Theorem 3.3.} Under the above two insecurity conditions, based on the proposed algorithm (with sorting), each sensor $i$ only needs to exclude $q_i$ suspicious sensors from its neighbouring set $\mathcal{N}_i$ to ensure the security of the entire network.

 \begin{proof}

	For sensor $i$, the effect of the detector (\ref{detector}) has been deduced in \cite{zhou2022security}, that is, $Pr\left( \gamma _{ij}=0 \right) =Pr( \| \hat{x}_i\left( k \right) -\hat{x}_{ij}^{a}\left( k \right) \| >\upsilon _i\xi _i\left( k \right) ) = 1-\exp ( -\upsilon _{i}^{-1}\left\| z_{ij}\left( k \right) \right\| )$,
where the  equation holds because the  parameter $\xi _i(k)$ satisfies that $\xi _i(k)\sim E\left( 1 \right) $, and  the values of $\hat{x}_i\left( k \right)$ and $\hat{x}_{j}\left( k \right)$ are very close.

For the insecurity condition in Definition 3.4, it is obvious that $Pr(y=0) \approx  1$, while for the insecurity condition in  Definition 3.5, the undetected attack signal $||z_{ij} ||$ is bounded with $||z_{ij}(k) ||\le\tilde{z}_i$ because $\upsilon_i^{-1}$ is set to a small positive number. 
Therefore, the augmented estimation error in  Definition 3.6 is bounded despite the presence of undetected attack signals.
That is, we have 
	\begin{equation}\label{error_bound}
		\left\| \lambda \left( \varUpsilon \left( k \right) \otimes A \right) Z\left( k \right) \right\| \le \lambda \max_i{\tilde{z}_i}\left\| A \right\| \sum_{i=1}^{|\mathcal{N}|}{|\mathcal{N}_i|},
	\end{equation}
if there exists a consensus parameter $\lambda$ such that $\rho(F(k))=\rho(\left(I_{\mathcal{N}}-\lambda \Gamma(k)\right) \otimes A-(\sum_{i=1}^{\mathcal{N}} \theta_{\mathcal{N}}^i \otimes K_i^{T}(k)) C)<1$ holds\cite{zhou2022security},
 where $\max_i{\tilde{z}_i}$ denotes the upper bound of undetected attacks among all $\mathcal{N}$ sensors, $i\in\mathcal{N}$.

 Combined with the above analysis,  the detector (\ref{detector}) is ultilized to verify the performance of the proposed algorithm  from different cases:
 \begin{enumerate}
     \item For both Definitions 3.4 and 3.5, for sensor $i$, if the $q_i$ suspicious sensors selected by the proposed algorithm are all detected to be attacked, it means that the $q_i$ attacked sensors  are correctly  excluded. In this case, the proposed algorithm can ensure the security of the entire network.
     \item For the unstealthy attacks in Definition 3.4, for sensor $i$, if only some of the $q_i$ suspicious sensors are detected to be attacked, it means that the proposed algorithm excludes more sensors on the basis of excluding all attack signals. However, according to Remark 2.1, the remaining $|\mathcal{N}_i|-q_i$ sensors can ensure that sensor $i$ achieves consistent decision-making. Therefore, the proposed algorithm can ensure the security of the entire system and also eliminate some larger noise.
     \item For the stealthy attack in Definition 3.5, for sensor $i$, if only some of the $q_i$ suspicious sensors are detected to be attacked, it does not mean that the remaining sensors are all safe. Therefore, the proposed algorithm excludes more sensors, which further limits the upper bound of  the attack signals that are not excluded and also eliminates some larger noise. And the relationship (\ref{error_bound}) shows that if the attack signals that are not excluded are upper bounded, then the proposed algorithm can guarantee the security of the entire system.
 \end{enumerate}

In summary, by applying the proposed algorithm to each sensor $i$, the security of the entire network can be guaranteed under two insecurity conditions. This completes the proof.
\end{proof}

\noindent {\bf Remark 3.3.} Theorem 3.3 proves that the proposed attack detection scheduling algorithm can guarantee that the augmented estimation error is bounded under two general unsecurity conditions. Moreover, it can be seen from \cite{zhou2022security} that the value of $\upsilon^{-1}$ has an impact on the detection accuracy. Therefore, in the simulation part, we will study the effect of different $\beta$ and $\upsilon^{-1}$ on the false positive rate and false negative rate.

\subsection{Discussion of the Effect of the Proposed Algorithm from Different Perspective}

\subsubsection{From the perspective of optimization rate}
~

Theorem 3.2 proves that for the entire time step, the theoretical lower bound of the average optimization rate of the proposed algorithm is $1-1/e$, which is higher than the average optimization rate proved to be $1/2$ in \cite{xu2022online}.
There are two factors that affect the optimization rate. One is the effect of noise uncertainty, that is, in some cases, attack signals (especially stealthy attacks) are indistinguishable from noise signals. The other is the effect of randomness, that is, the selection of sensors according to the vector of distribution proportions $p_k^{(l)}$ is random, but the randomness can be eliminated by sorting.

\subsubsection{From the perspective of real-time performance}
~

Attack detection is a task with high real-time requirements. However, whether the time taken for suspicious sensorss selection will affect the real-time performance of attack detection should be considered. The literature \cite{wang2018vision} shows that the data update frequency of the sensor is not infinite. For example, the update frequency of the image sensor is generally $10Hz-30Hz$, and the update frequency of the inertial sensor is $100Hz-1kHz$. So we only need to complete the detection task within the time interval between two sensor data updates.
 With the advancement of computer technology, the computering power has been significantly improved, so the proposed attack detection scheduling algorithm can be realized without affecting the real-time performance.

\subsubsection{From the perspective of complexity}
~

Regarding the complexity of the proposed algorithm, the complexity of Algorithm 1 is $\mathcal{O}(q_i|\mathcal{N}_i|)$. And if the sorting is completed before each selection and then take the maximum value, the complexity will be updated to $\mathcal{O}(q_i|\mathcal{N}_i|log(|\mathcal{N}_i|)$.
However, Yang et al. \cite{yang2021secure} directly selected the $q_i$ optimal sensors from $\mathcal{N}_i$ with a complexity of $\mathcal{O}(q_i\binom{|\mathcal{N}_i|}{q_i})$, which is difficult to achieve for larger $|\mathcal{N}_i|$ in large-scale networks. 
At this time, the complexity of the proposed algorithm is still lower than that of the existing algorithm.

\subsubsection{From the perspective of energy and computing power requirments}
~

The attack detection is a module deployed on each sensor to judge the security of the data transmitted by the surrounding sensors, which essentially uses some algorithms and technologies for data analysis and has certain computing power requirements. Therefore, the complexity of different algorithms deployed on the attack detection module is an important factor affecting the computing power requirements. Similar to the existing literature on sensor energy consumption \cite{wang2021sleep}, the energy consumption of attack detection is related to  the power of calling the attack detection module $P_{detect}$ and the time of calling the module (running the algorithm) $t_d$.   Similarly, we also believe that the power of calling the attack detection module is a constant value. That is, the longer the time to call the attack detection module, the higher the energy consumption of attack detection. Theorem 3.3 shows that based on the proposed algorithm, each sensor can directly  search for the set of suspicious sensors to ensure the security of the entire network. Therefore, compared with the existing algorithm \cite{guan2017distributed,zhou2022security,yang2019distributed}, the proposed algorithm no longer needs to detect information from all neighbor sensors, which can reduce the time of calling the attack detection module.

\subsubsection{From the perspective of universality}
~

The proposed algorithm is general and can be applied to other possible network topologies, network scales, and attacker settings. For distributed networks with different topologies, the proposed algorithm can be applied as long as the network topology satisfies the undirected graph structure defined in Section IIA. The proposed algorithm can be applied to networks with different scales, not limited to large-scale or small-scale networks. It should be pointed out that the larger the scale of the network, the less energy consumed by the proposed algorithm to directly select suspicious sensors compared to the existing literature to detect information from all sensors. For attacker settings, Definition 3.4 and 3.5 give general definitions of unstealthy attacks and stealthy attacks, that is, unstealthy attacks and stealthy attacks in existing literature generally follow these two definitions. For example, unstealthy attacks in \cite{basiri2019kalman} and stealthy attacks in \cite{song2022optimal,guo2018worst}. Therefore, the proposed algorithm is also feasible for different attack settings.

\section{Simulation and Experimental Results}
In this section, the effectiveness of the proposed method is jointly validated  by numerical simulation and practical experimental results. First, numerical simulation results verify the performance of the proposed algorithm. Then, as a supplementary experiment, the practical experiment illustrate the advantages of the proposed algorithm in terms of energy consumption.
\subsection{Numerical Simulation}
\begin{figure}[!h]
	\centering
	\includegraphics[width=1.8in]{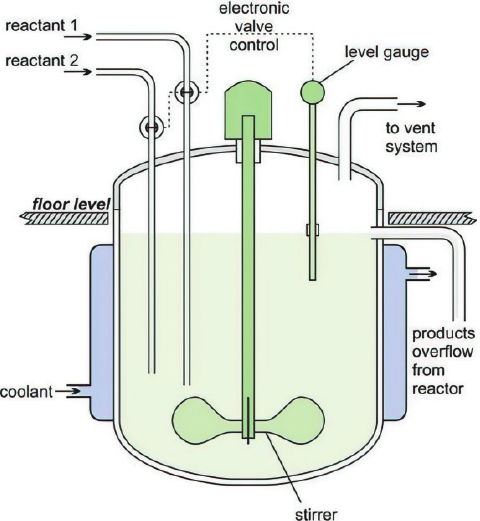}
	\caption{A physical structure of a continuous stirred tank		reactor (CSTR)}
	\label{fig1}
\end{figure}

To demonstrate the effectiveness of the proposed algorithm, we consider a real scenario, an industrial continuous stirred tank reactor (CSTR), where the output concentration of the educt and the reactor temperature are required to be monitored, as shown in Fig.\ref{fig1} \cite{guan2017distributed}. 
\begin{figure}[!h]
	\centering
	\includegraphics[width=3in]{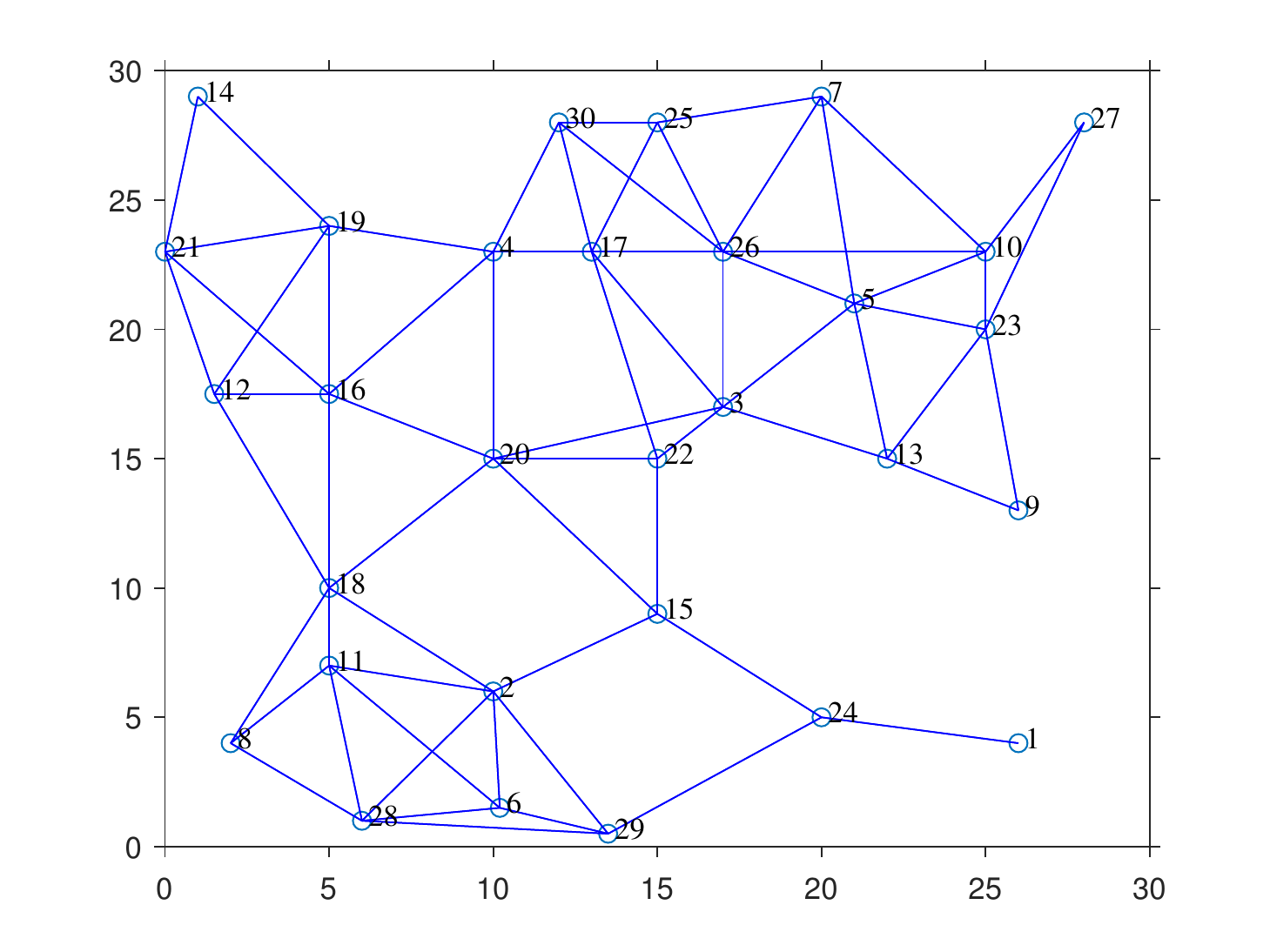}
	\caption{Topology of the sensor network in numerical simulation.}
	\label{fig2}
\end{figure}
The system model is as follows
\begin{eqnarray} 
	x\left( k+1 \right) =Ax\left( k \right) +\omega \left( k \right) ,
\end{eqnarray}
where
\begin{eqnarray} \label{matrix_A}
	A=\left[ \begin{matrix}
	0.9719&		-0.0013\\
	-0.0340&		0.8628\\
	\end{matrix} \right] ,
\end{eqnarray}
and $x\left( k \right) =\left[ C_A\left( k \right), \,\,T  \right]^T $, where $C_A$ is the output concentration of the educt $A$, and $T$ denotes the reactor temperature. 
\begin{figure*}
\centering
\begin{minipage}[!b]{\linewidth}
\subfigure[$\beta=0.2$]{
 			\includegraphics[width=0.31\linewidth]{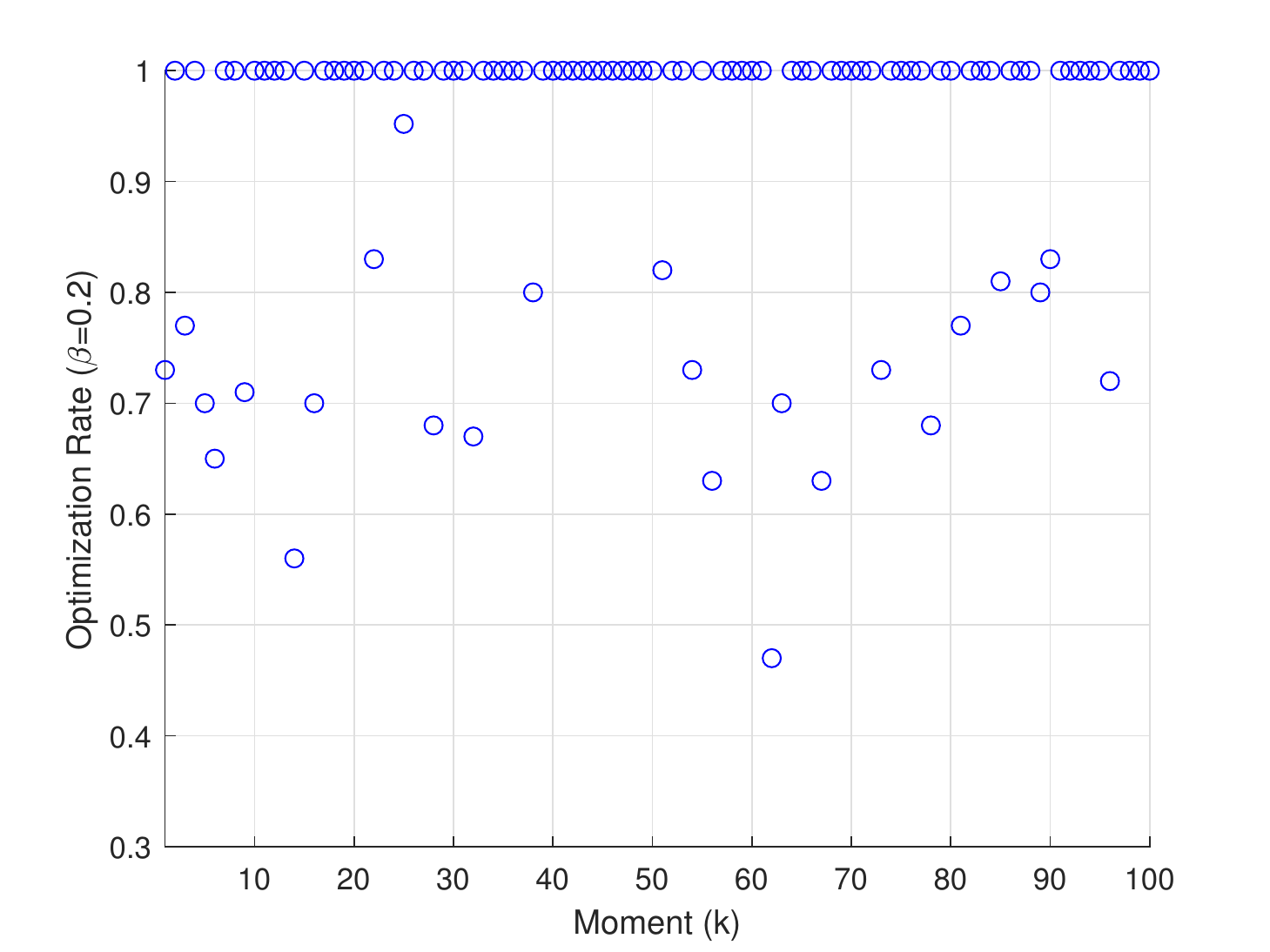}
 		}
 		\subfigure[$\beta=0.5$]{
 			\includegraphics[width=0.31\linewidth]{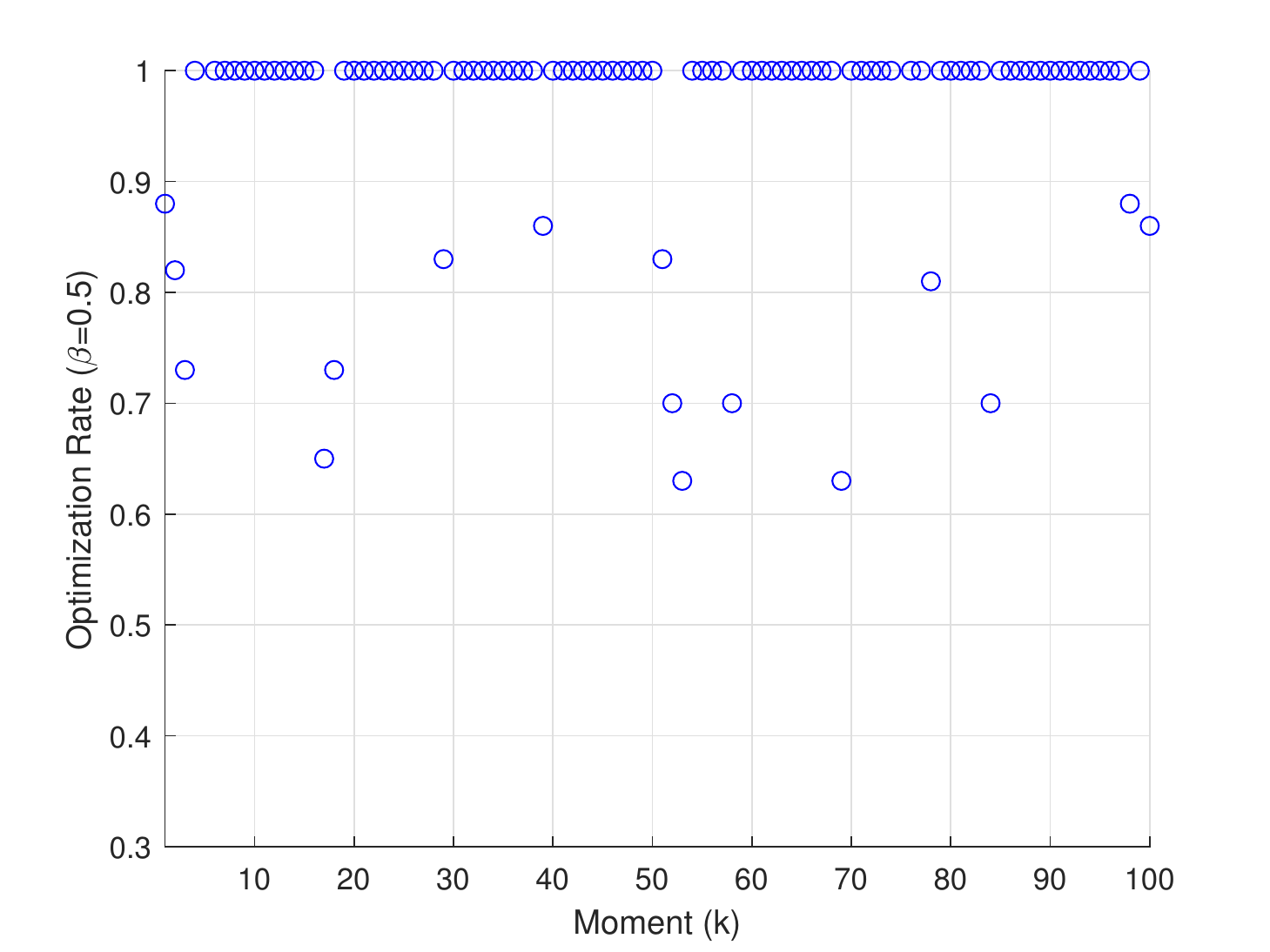}
 		}
 		\subfigure[$\beta=1$]{
 			\includegraphics[width=0.31\linewidth]{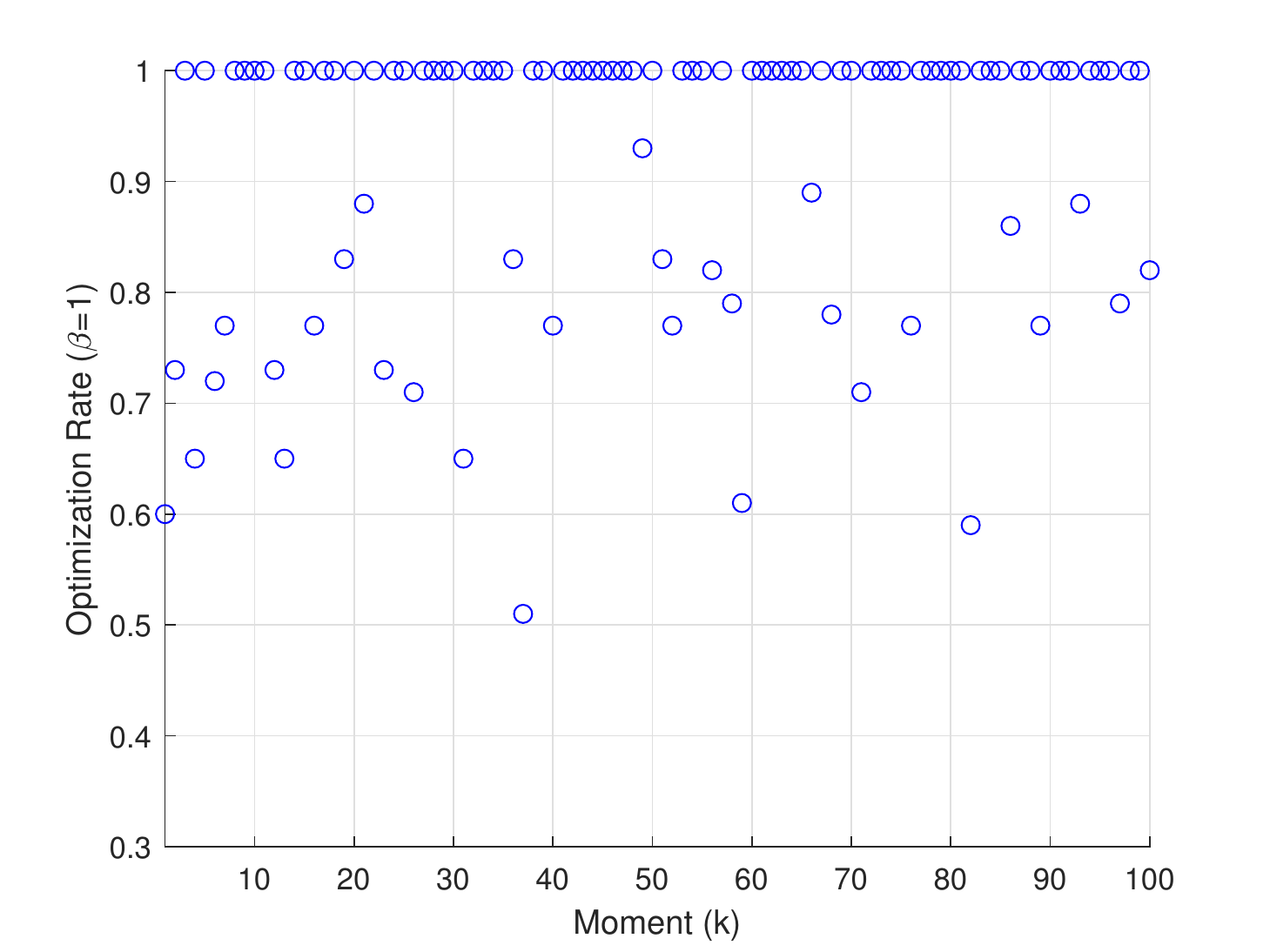}
 		}
\caption{The optimization rate under unstealthy attacks with different $\beta$.}\label{fig3}
\end{minipage}
\end{figure*}
\begin{figure*}
\centering
\begin{minipage}[!h]{\linewidth}
\subfigure[$\beta=0.2$]{
 			\includegraphics[width=0.31\linewidth]{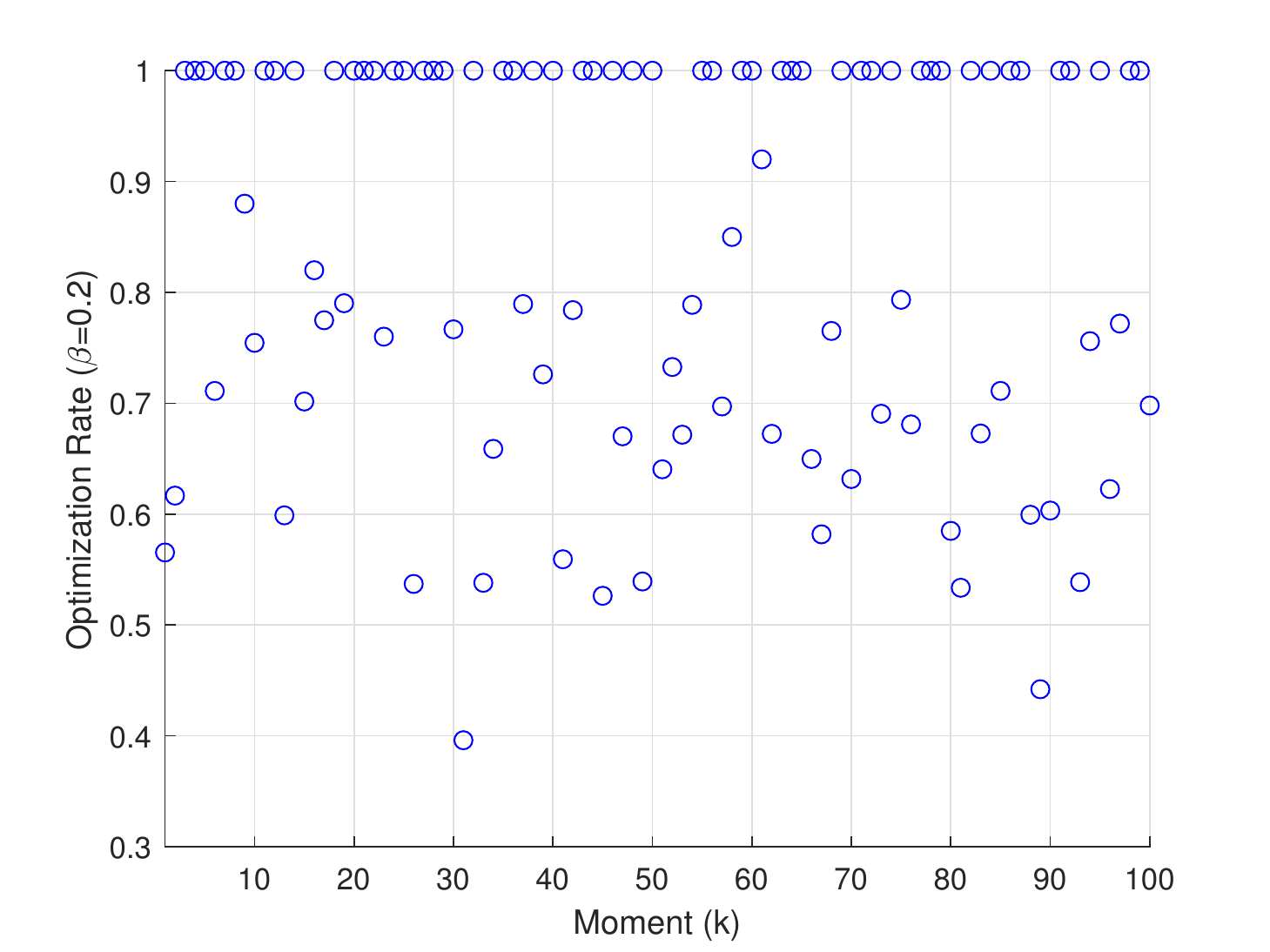}
 		}
 		\subfigure[$\beta=0.5$]{
 			\includegraphics[width=0.31\linewidth]{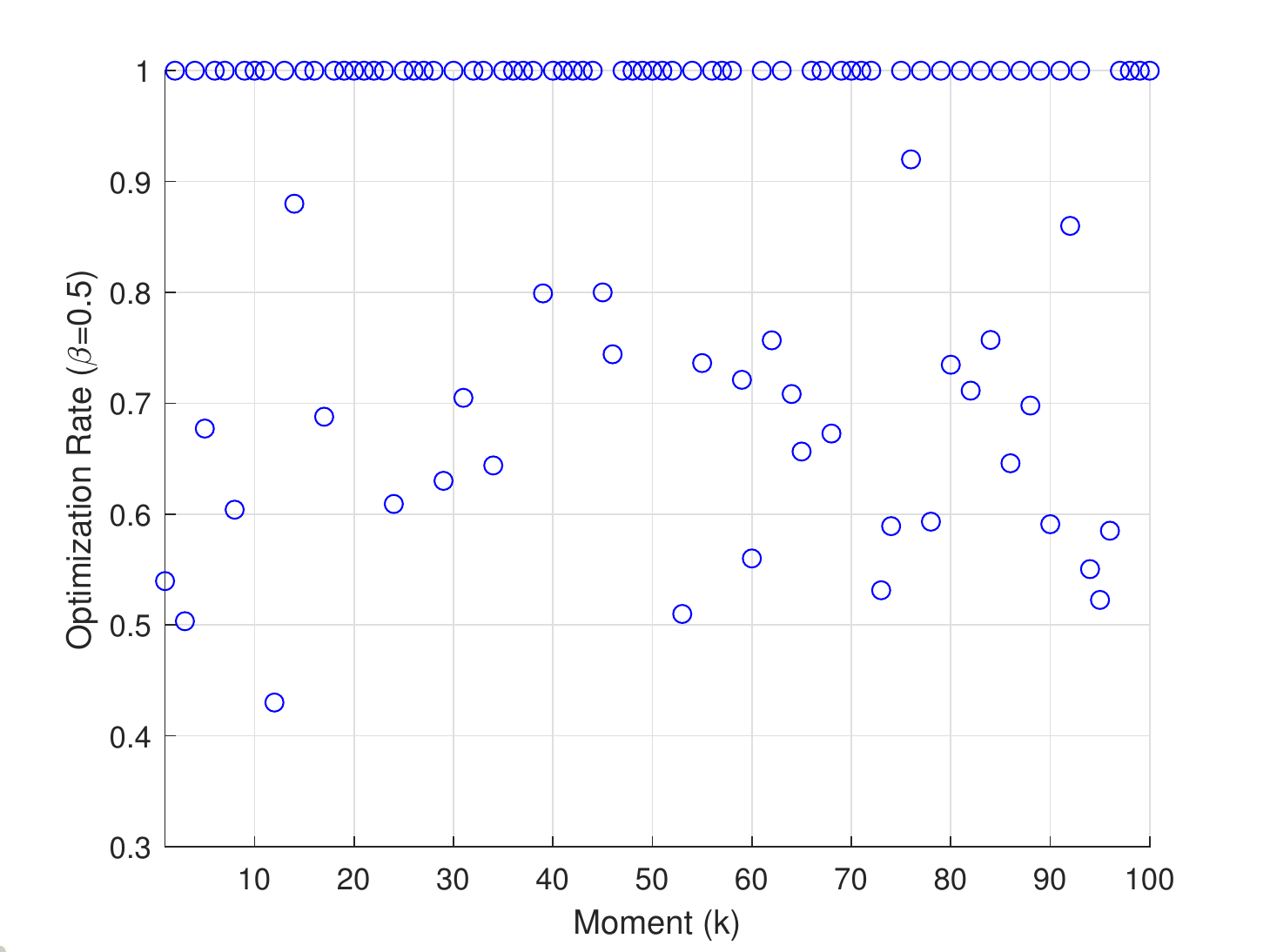}
 		}
 		\subfigure[$\beta=1$]{
 			\includegraphics[width=0.31\linewidth]{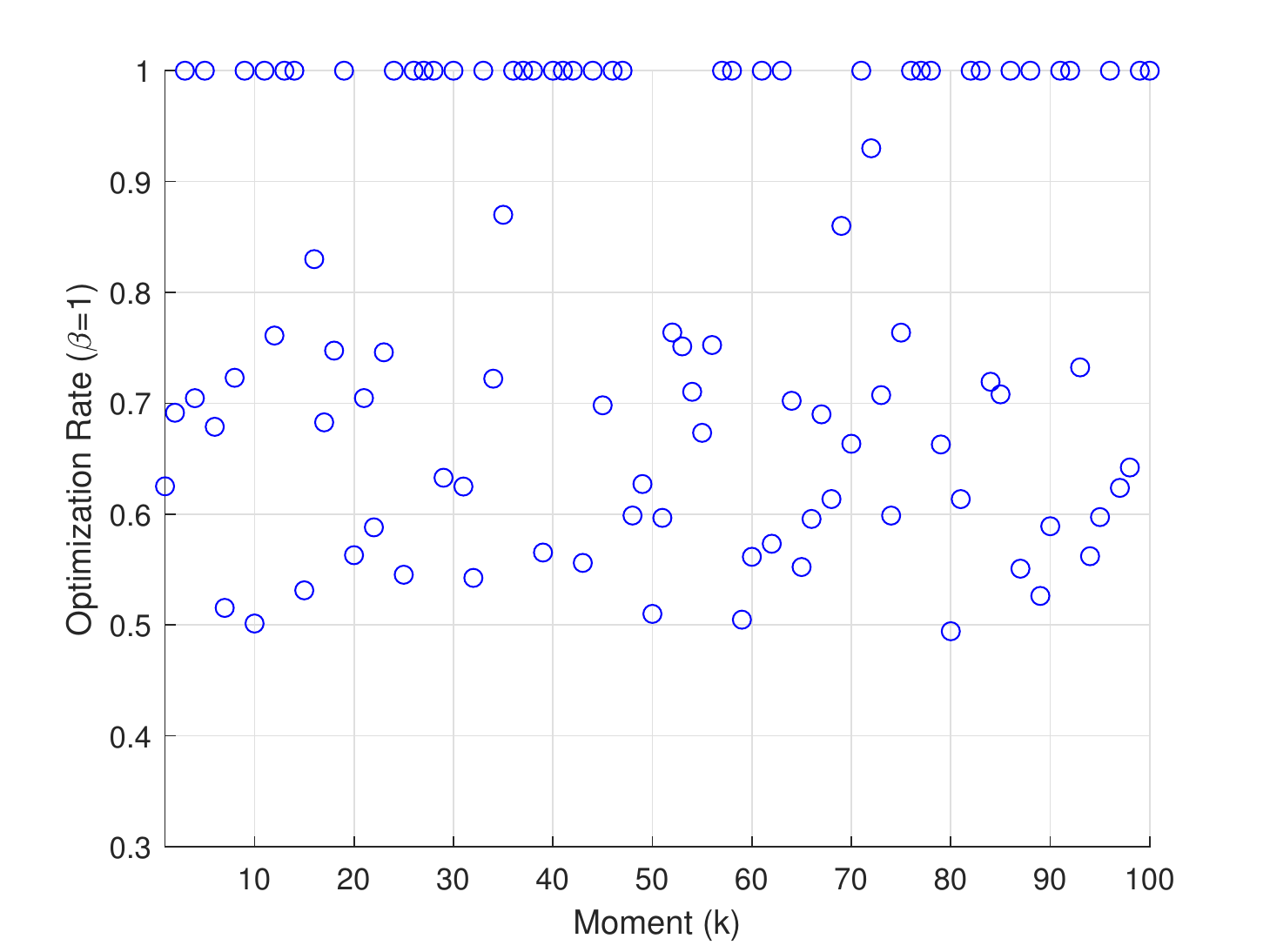}
 		}
\caption{The optimization rate under stealthy attacks with different $\beta$.}\label{fig4}
\end{minipage}
\end{figure*}

We consider deploying a distributed sensor network as shown in Fig. \ref{fig2} to monitor the states of the system. There are 30 sensors distributed in a space of size $30 units \times 30 units$. For sensor $i$, $i\in \mathcal{N}$, the measurement model is
\begin{eqnarray} 
	y_{i}\left( k \right) =C_{i}x\left( k \right) +\nu _{i}\left( k \right) ,
\end{eqnarray}
where $C_i=\left[ 0, \,\,0.1+1/i  \right]^T $.
The system parameters are defined as follows: $Q=0.5I$, $R_i=0.5I$, $\lambda=0.1$, $\beta\in[0,1]$. Moreover, both $||\omega(k)||$ and $||\nu_{i}(k)||$ are upper bounded by $0.05$. The considered attack signals include two types, one of which is unstealthy attacks with large amplitude at most times in Definition 3.4, and the other is stealthy attacks with amplitude close to noise at most times in Definition 3.5.

Based on the above real scenario, the following two cases are given to illustrate the effectiveness of the algorithm in this paper.

\textbf{Case 1:} the case where only some of the sensors (sensor 5 and the surrounding sensors 3, 7, 10, 13, 23, 26) in Fig. \ref{fig2} are considered.

According to the Assumption 2.1, the maximum number of attacked neighboring sensors around sensor $i$  is $\lfloor \mathcal{N}_i/2 \rfloor=3$. It is assumed that the attacker launches FDIAs from the initial moment $k=1$, and dynamically adjusts the attack strategy. At $k = [1, 50]$, communication links (5,7), (5,10), (5,23) are attacked, $k=[51, 100]$, communication links (5,3), (5,7), (5,23) are attacked.

To verify the effectiveness of Theorem 3.2, we explore the optimization rate of the proposed algorithm for the two attacks at various moments under different $\beta$ (the optimization rate is not related to the accuracy of the detector, but only related to the selection of the set of suspicious sensors and optimal set of suspicious sensors).
Take the optimization rate in Definition 3.3 as the evaluation metric. Obviously, the closer the value of the optimization rate is to $1$, the more accurate the selection of suspicious sensors is. Fig. \ref{fig3} and Fig. \ref{fig4} show the optimization rate when $\beta$ is different under unstealthy and stealthy attacks, respectively (This simulation sorts the vector of distribution proportions before selecting a sensor). 
For unstealthy attacks, the optimization rate of the proposed algorithm has little difference under different values of $\beta$. For stealthy attacks, the optimization rate is better when $\beta=0.5$ than when $\beta$ is $0.2$ or $1$, because the proposed algorithm uses historical information to guide the selection of suspicious sensors at the current moment. Therefore, the coefficient $\beta$ in subsequent simulations is taken as $0.5$ if not otherwise specified to better weigh the historical information and the current information.

Then, take the average optimization rate in Definition 3.3 as the evaluation metric (according to Theorem 3.2, the theoretical lower bound of the average optimization rate is $1-1/e$). To demonstrate the performance of the proposed algorithm, we compare the average optimization rate of four algorithms under unstealthy attacks and stealthy attacks, as shown in Table \ref{tab2}. The four algorithms  are the algorithm in \cite{xu2022online}, the proposed Algorithm $1$ (without sorting), the algorithm in \cite{yang2021secure}, and the proposed Algorithm $1$ (with sorting).
 It can be seen that the average optimization rate of the proposed algorithm (without sorting) is higher than that of the algorithm in \cite{xu2022online}. The average optimization rate of the proposed algorithm (with sorting) is almost the same as the algorithm in \cite{yang2021secure}, but the complexity of the proposed algorithm is lower, 
 that is, $\mathcal{O}(q_i|\mathcal{N}_i|log(|\mathcal{N}_i|)$ is lower than $\mathcal{O}(q_i\binom{|\mathcal{N}_i|}{q_i})$. In addition, all algorithms perform better under unstealthy attacks than under stealthy attacks, because unstealthy attacks are easier to distinguish from noise signals.

\begin{table}[!h]
 \caption{The average optimization rate of different algorithms under unstealthy and stealthy attacks.}
	\label{tab2}

  \begin{center}
\begin{tabular}{|l||l|l|}
\hline
Index                                                                                & \multicolumn{2}{l|}{Average Optimization Rate}                       \\ \hline
Attacks Types                                                                        & \multicolumn{1}{l|}{Unstealthy Attacks}     & Stealthy Attacks       \\ \hline
\multirow{2}{*}{The algorithm in {[}36{]}}                                           & \multicolumn{1}{l|}{\multirow{2}{*}{0.650}} & \multirow{2}{*}{0.576} \\
                                                                                     & \multicolumn{1}{l|}{}                       &                        \\ \hline
\begin{tabular}[c]{@{}l@{}}The proposed Algorithm 1\\ (without sorting)\end{tabular} & \multicolumn{1}{l|}{0.744}                  & 0.661                  \\ \hline
\multirow{2}{*}{The algorithm in {[}32{]}}                                           & \multicolumn{1}{l|}{\multirow{2}{*}{0.951}} & \multirow{2}{*}{0.873} \\
                                                                                     & \multicolumn{1}{l|}{}                       &                        \\ \hline
\begin{tabular}[c]{@{}l@{}}The proposed Algorithm 1\\ (with sorting)\end{tabular}    & \multicolumn{1}{l|}{0.946}                  & 0.870                  \\ \hline
\end{tabular}\\
\end{center}

\end{table}

\begin{figure}
\centering
\begin{minipage}[!b]{\linewidth}
\subfigure[False Negetive Rate]{
 			\includegraphics[width=0.46\linewidth]{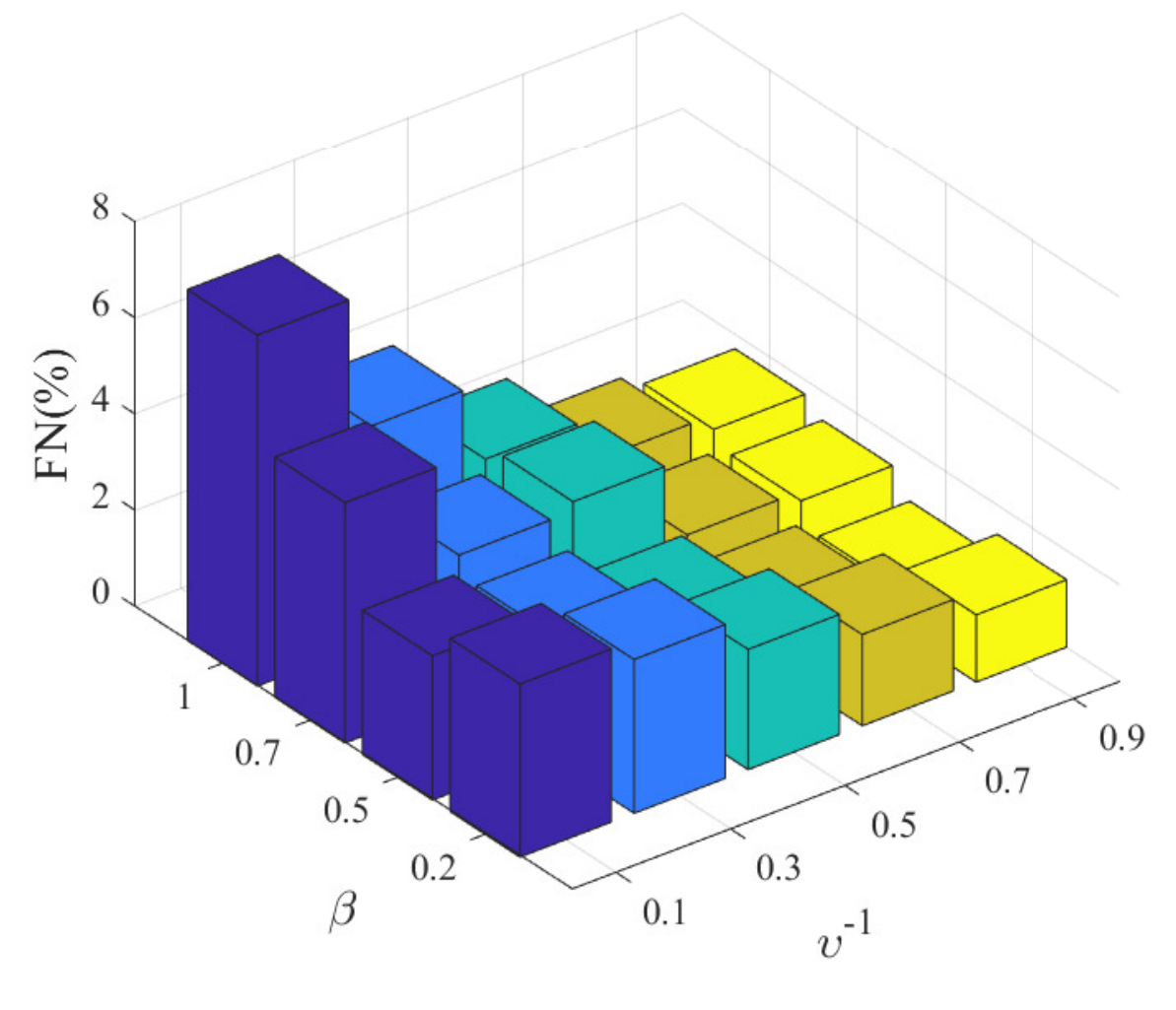}
 		}
 		\subfigure[False Positive Rate]{
 			\includegraphics[width=0.46\linewidth]{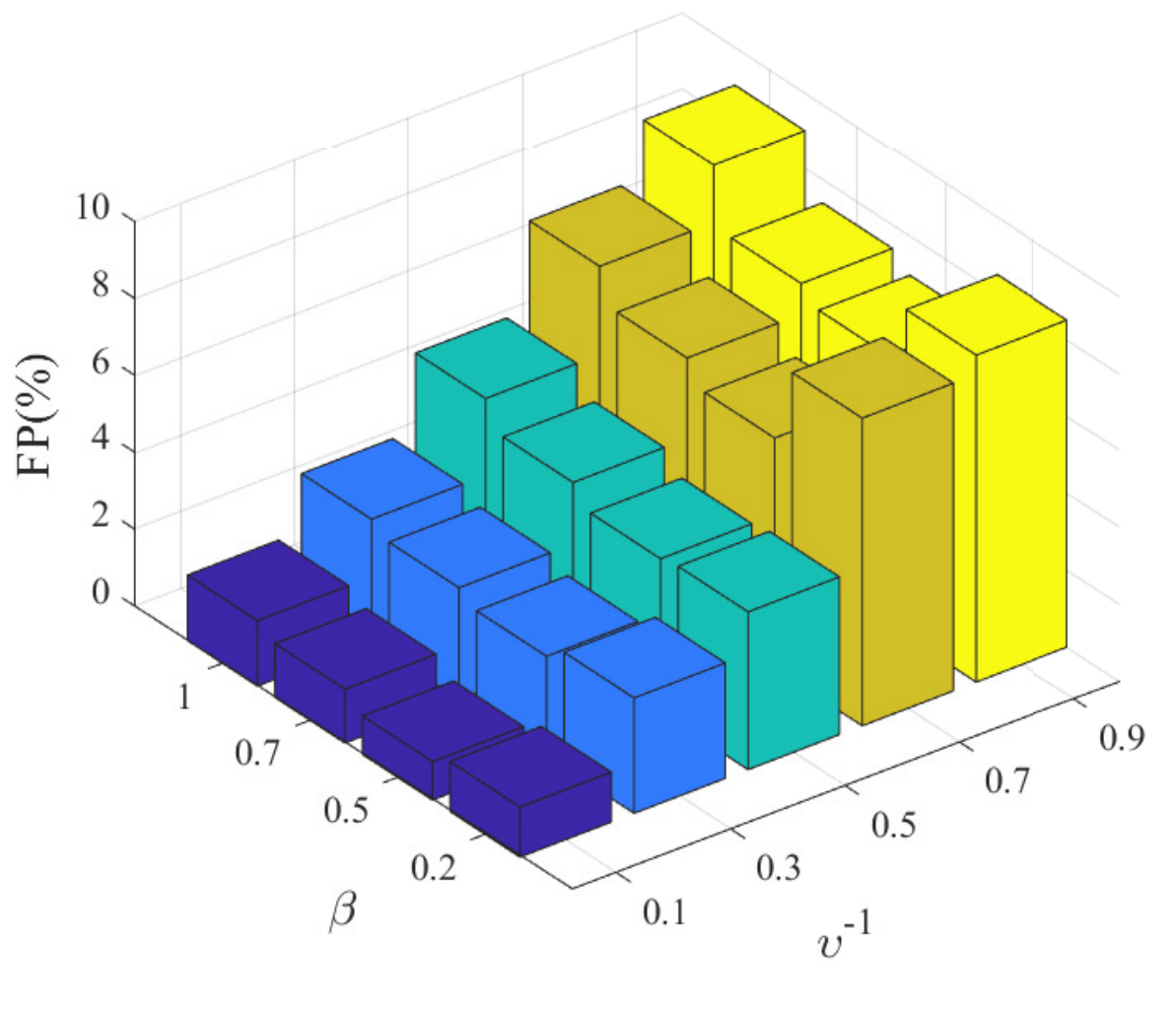}
 		}
\caption{The relation of $\upsilon^{-1}$
		to false negative rate and false positive rate with different $\beta$ under stealthy
		attacks.}\label{fig6}
\end{minipage}
\end{figure}

Since stealthy attacks are more difficult to detect than unstealthy attacks, we take the stealthy attacks as an example to simulate the attack detection effect of the proposed algorithm (with sorting) with different detection threshold $\upsilon^{-1}$ and coefficient $\beta$. Take the False Negative Rate (FN) and False Positive Rate (FP) as the evaluation metrics.
It can be seen from the Fig. \ref{fig6} that as the threshold $\upsilon^{-1}$ increases, FN becomes lower and FP becomes larger. Also, the appropriate values of $\beta$ and $\upsilon^{-1}$ allow both FN and FP to be low, i.e., when $\beta$ and $\upsilon^{-1}$ both takes $0.5$. The proposed algorithm (with sorting) has lower FN and FP than the detector in \cite{zhou2022security} because we directly select the suspicious sensors.

\textbf{Case 2:} the case where the entire network as shown in Fig. 
\ref{fig2} is considered.

The simulation results in \textbf{Case 1} show that the proposed Algorithm 1 is effective in detecting unstealthy attacks, so we only need to verify whether the estimation error is bounded under stealthy attacks to prove the effect of Theorem 3.3.

Suppose the attacker starts to launch FDIAs from the moment $k=100$. That is, at $k=[101,500]$, communication links (2,15), (2,29), (5,7), (5,10), (5,23), (16,12) and (16,19) are attacked. 
 
\begin{figure}[!b]
	\centering
	\includegraphics[width=3.5in]{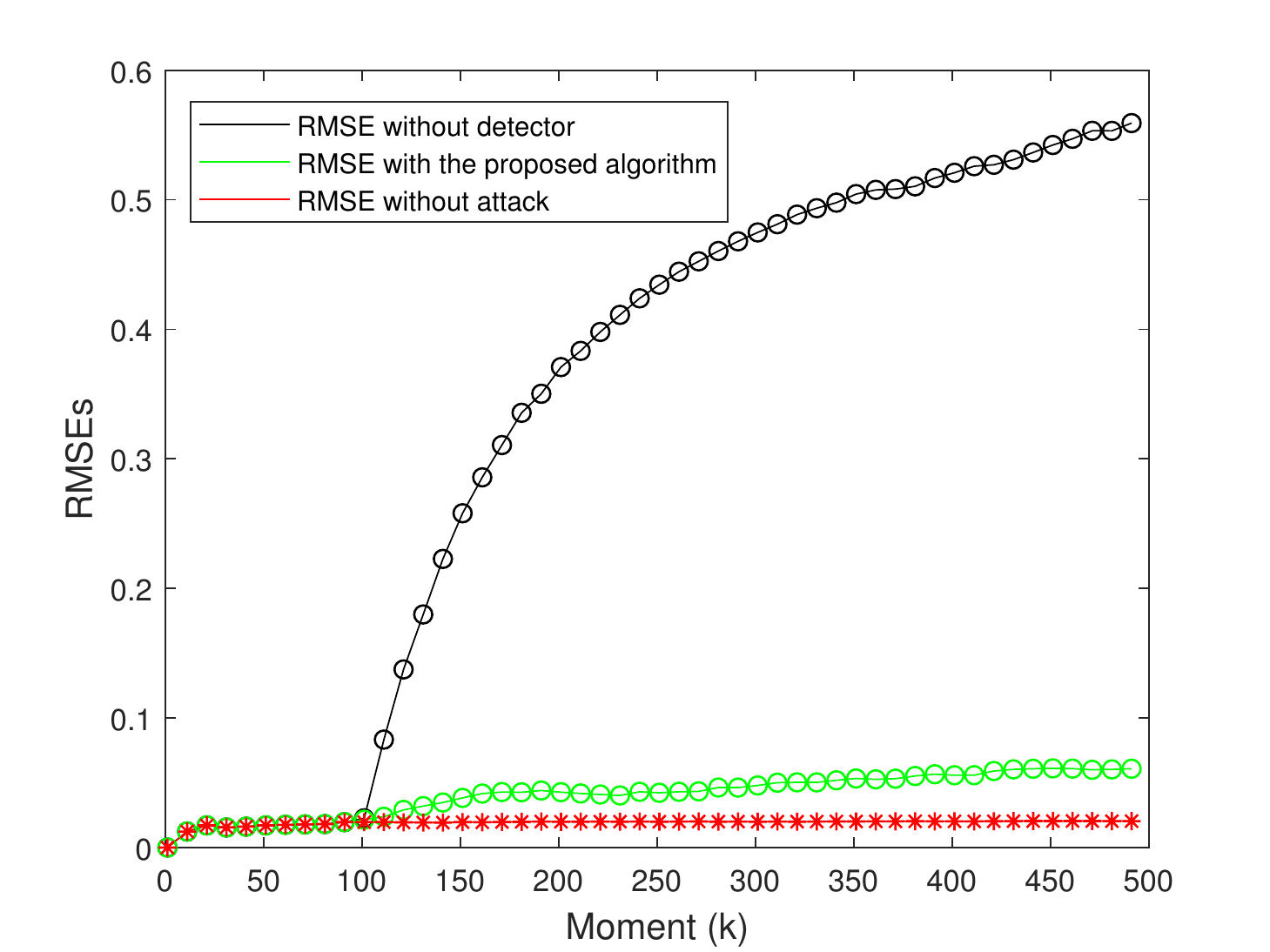}
	\caption{Comparative experiment of RMSEs under different algorithms.}
	\label{fig8}
\end{figure}

 The estimation error evaluation metric used in this paper is the root mean square error (RMSE), i.e., $RMSE\left( k \right) =\sqrt{\frac{1}{Z}\sum_{i=1}^Z{||e(k,z)||^2}}$, where $Z$ denotes the number of Monte Carlo experiments and $||e(k,i)||$ denotes the norm of the average estimation error at moment $k$ in the $i$-th Monte Carlo experiment.
 Fig. \ref{fig8} compares the RMSE under different algorithms, including the RMSE without detector, the RMSE with the proposed algorithm,  and the RMSE without attack. It can be seen that the algorithm proposed in this paper  can guarantee that the RMSEs are bounded (slightly higher than the RMSE without attack) which verifies the conclusion of Theorem 3.3.  
 It should be noted that with the aid of the proposed algorithm, each sensor $i$ can directly find the set of suspicious sensors from its neighbor sensors, which can reduce the energy consumption of attack detection. The following practical experiment will further verify the advantages of the proposed algorithm in terms of energy consumption.

\subsection{Practical Experiment}
\begin{figure}[!b]
	\centering
	\includegraphics[width=3in]{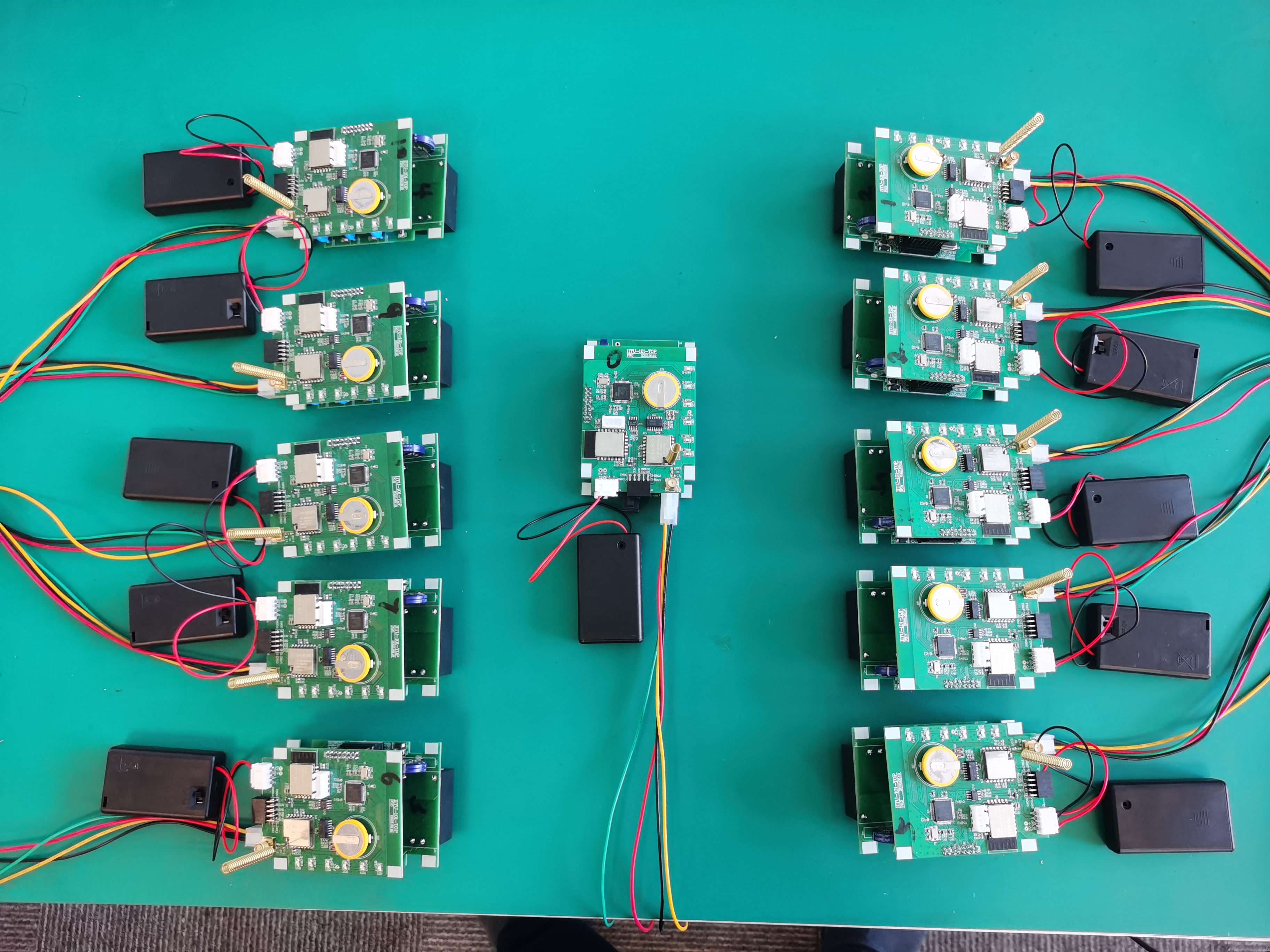}
	\caption{Topology of the sensor network in experimental environment.}
	\label{physical_structure_diagram}
\end{figure}

To verify the advantages of the proposed algorithm in terms of energy consumption, we deployed the proposed algorithm in an experimental environment. The topology of the sensor network in the experimental environment is shown in Fig. \ref{physical_structure_diagram}, including a central sensor (ID: 0) and $10$ sensors (ID: 1-10). The system to be monitored is a three-phase electrical system. The yellow, green, and red wires in Fig. \ref{physical_structure_diagram} correspond to the three-phase($A$, $B$, and $C$)  electricity drawn from the system. 
In Fig. \ref{physical_structure_diagram}, each sensor is controlled by a GD32 microcontroller, and the wireless communication between each sensor and the central sensor is based on the Lora module. Each sensor is powered by three AAA batteries connected in series. Therefore, all sensors in this experiment are energy limited.

The state variables are the effective value of the three-phase voltage of the power supply system, which is defined as $x(k)=[V_A(k),V_B(k),V_C(k)]^T$. All sensors (including the central sensor) can measure the values of state variables. At each moment, the central sensor not only measures the system, but also receives measurements from neighboring sensors.
However, the measurement results may be tampered with by malicious adversaries during the transmission process. Assume  that $q_i=2$ sensors out of the $10$ sensors are attacked (the number is less than $5$). Therefore, the proposed algorithm can be applied on the central sensor to exclude suspicious sensors, thereby ensuring the security of state estimation. 

\begin{figure}[!h]
	\centering
	\includegraphics[width=3.4in]{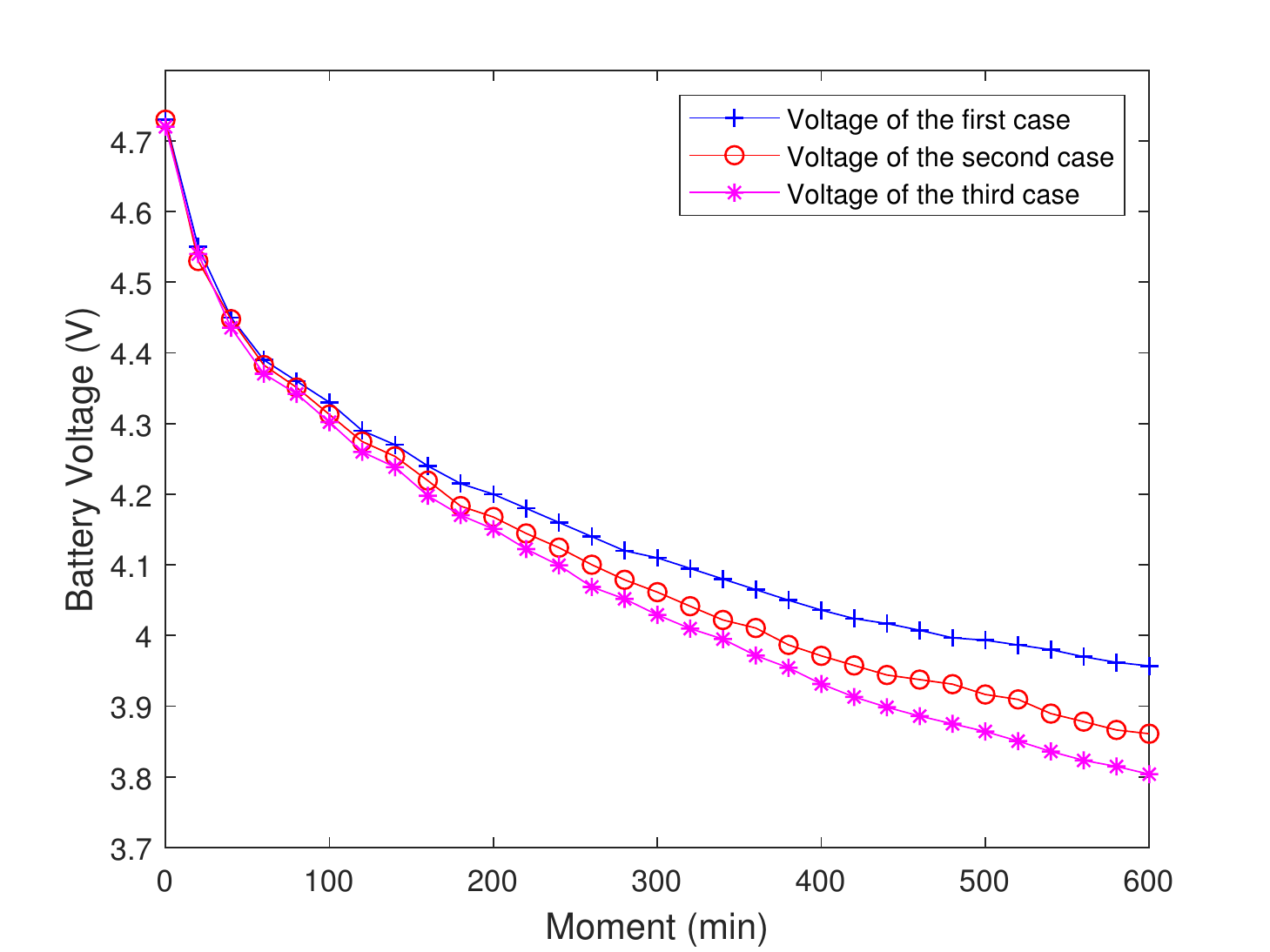}
	\caption{The voltage change curves of the central sensor in different cases.}
	\label{fig_energy}
\end{figure}

This experiment focuses on the energy consumption of the central sensor, which can be reflected by the voltage change of the battery. In different cases, we measured the voltage change of the center sensor over a period of time, as shown in Fig. \ref{fig_energy}. The first case  is the voltage change curve (blue) without deploying any attack detection algorithm, which is called the base curve (the energy consumption mainly comes from data transmission and other basic energy consumption). The second case  is the voltage change curve (red) when the attack detection scheduling algorithm ($\beta=0.5$) and the attack detection algorithm are deployed at the same time. The third case  is the voltage change curve (mulberry) when only the attack detection algorithm is deployed.

Preliminarily, it can be seen from Fig. \ref{fig_energy} that the voltage change curve in the second case is higher than the voltage change curve in the third case, which means that the proposed algorithm has lower energy consumption.
Furthermore, the initial voltages of these three cases are slightly different, we use the magnitude of the voltage drop in the three cases within the same time period to analyze the level of energy consumption. At $600$ min, the voltage in the first case dropped by about $0.773V$ (base value), and the voltages in the second and third cases dropped by about $0.871V$ and $0.918V$, respectively. Approximately, the energy consumption of the second case is  $1-(0.871-0.773)/(0.918-0.773)\approx 32.4\%$ lower than that of the third case.  In summary, compared with the traditional attack detection algorithm, the proposed algorithm can reduce energy consumption by directly selecting suspicious sensors.

\section{Conclusions}
This paper considers the scheduling problem of attack detection on large-scale networks under FDIAs. First, we transform the NP-hard suspicious sensor set selection problem into a solvable submodular problem. Then, we propose an attack detection scheduling algorithm based on the sequential submodular optimization theory, which can guarantee a theoretical lower bound on the average optimization rate. Finally, it is theoretically proved that the proposed algorithm can ensure that the augmented estimation error of the entire network is bounded. In the future, we will consider how to improve the efficiency of suspicious sensor selection to further reduce energy consumption.

\vfill

\end{document}